\documentclass[pra,superscriptaddress,amsmath,%
amsfonts,amssymb,a4paper,twocolumn]{revtex4}

\usepackage[english]{babel}
\usepackage{epsfig}

\def\be{\begin{equation}}
\def\ee{\end{equation}}
\def\bea{\begin{eqnarray}}
\def\eea{\end{eqnarray}}

\def\a{\alpha}
\def\b{\beta}
\def\d{\delta}

\begin{document}


\title {Exactly-solvable coupled-channel potential models of atom-atom
magnetic Feshbach resonances from supersymmetric quantum mechanics}

\author{Andrey M. Pupasov}
\affiliation{Physics Department, Tomsk State University, 36 Lenin Avenue,
634050 Tomsk, Russia}

\author{Boris F.\ Samsonov}
\affiliation{Physics Department, Tomsk State University, 36 Lenin Avenue,
634050 Tomsk, Russia}

\author{Jean-Marc Sparenberg}
\affiliation{Physique Quantique, C.P.\ 229, Universit\'e Libre de Bruxelles,
B 1050 Bruxelles, Belgium}

\date{\today}

\begin{abstract}
Starting from a system of $N$ radial Schr\"odinger equations with a vanishing potential and finite threshold differences between the channels,
 a coupled $N \times N$ exactly-solvable potential model is obtained with the help of a single non-conservative supersymmetric transformation.
The obtained potential matrix, which subsumes a result obtained in the literature, has a compact analytical form, as well as its Jost matrix.
It depends on $N (N+1)/2$ unconstrained parameters and on one upper-bounded parameter, the factorization energy.
A detailed study of the model is done for the $2\times 2$ case:
a geometrical analysis of the zeros of the Jost-matrix determinant shows that the model has 0, 1 or 2 bound states, and 0 or 1 resonance;
the potential parameters are explicitly expressed in terms of its bound-state energies, of its resonance energy and width, or of the
open-channel scattering length, which solves schematic inverse problems.
As a first physical application, exactly-solvable $2\times 2$ atom-atom interaction potentials are constructed, for cases where
a magnetic Feshbach resonance interplays with a bound or virtual state close to threshold, which results in a large background scattering length.
\end{abstract}

\pacs{03.65.Nk, 24.10.Eq, 34.20.Cf, 34.50.-s}

\maketitle


\section{Introduction}

Coupled-channel quantum-scattering models are experiencing today a
strong renewal of interest, mainly thanks to the impressive
experimental progress in the field of ultracold gases. Actually,
an indispensable tool to control the atom-atom interactions in
these systems relies on the coupling between channels defined by
different hyperfine states of the atom-atom pairs. When dipped in
a magnetic field, these hyperfine states have threshold energies
which vary linearly with the field. At ultracold temperatures, the
lowest-threshold state is the only open channel, while states with
higher thresholds are closed. When varying the field in a
well-chosen range, quasi bound states of the closed channels can
appear as resonances in the open channel, a phenomenon known as a
``Feshbach resonance'', a phenomenon first studied in nuclear
physics~\cite{feshbach:58,feshbach:62}. When a Feshbach resonance
crosses the open-channel threshold, due to magnetic-field
variation, the scattering length $a$, which effectively controls
the atom-atom interaction, goes through infinite values, switching
from positive to negative
sign~\cite{tiesinga:92,tiesinga:93,moerdijk:95}. This very
spectacular phenomenon is now known as a magnetic-field-induced
Feshbach resonance, or just ``magnetic Feshbach resonance''.

The practical importance of magnetic Feshbach resonances has motivated various theoretical models, which can be classified in three categories:
(i) microscopic models, which should in principle deduce magnetic-Feshbach-resonance properties from many-electron calculations;
(ii) effective potential models, which reduce the complexity of the many-electron problem to a two-atom coupled-channel problem (usually two channels are enough), where the interaction between the two atoms is modeled by a symmetric (two by two for a two-channel model) potential matrix;
(iii) effective scattering-matrix models, which reduce the role of the underlying interactions to its impact on the atom-atom open-channel scattering matrix.
Present-day theoretical descriptions of ultracold gases only require the knowledge of the atom-atom scattering length, which is directly related to the open-channel scattering matrix.
As far as practical applications are concerned, the third category of models is thus sufficient.
There, a magnetic Feshbach resonance is described as a pole of the scattering matrix in the complex wave-number planes, like any resonance~\cite{taylor:72},
and the whole complexity of the many-electron or atom-atom problem is reduced to a few parameters of the scattering-matrix Pad\'e expansion~\cite{marcelis:04}.
These parameters can be numerically obtained, e.g., with the help of the reaction-matrix method~\cite{nygaard:06}, from a given microscopic or effective-potential model.

In several contexts, the use of effective-scattering-matrix models is however progressively felt to be insufficient. This is for instance the case when a magnetic Feshbach resonance occurs with a large background scattering length, due to a bound or virtual state in the open channel, close to its threshold~\cite{marcelis:04}. Such an open-channel state is also called a ``potential resonance'' because it naturally occurs in a potential model, even in a single-channel case. Other situations where a more detailed knowledge of the atom-atom interaction than just the scattering length might be necessary are cases where molecules can be formed, as in crossovers between a Bardeen-Cooper-Schieffer superfluid and a Bose-Einstein condensate, or in Bose-Einstein-condensate collapses. None of these cases probably requires to solve the full many-body electronic problem; effective-potential models, on the other hand, look like a reasonable approximation, as they allow for a realistic description of the atom-atom interaction in terms of the accessible channels and as a function of the radial coordinate $r$ between atoms.

There is thus an interest for exactly-solvable coupled-channel
potential models with threshold differences. The first example
coming to mind is probably the coupled square-well potential,
which can display both potential and Feshbach resonances, as well
as bound states~\cite{kokkelmans:02}. This model has however two
drawbacks: first, despite its simplicity and exactly-solvable
character, its scattering-matrix poles are given by rather
complicated implicit equations. Second, its discontinuous form
factor is rather limitative and very different from the known
long-range atom-atom polarization interaction. The next choice
towards realistic atom-atom interactions is thus a purely
numerical resolution of the coupled-channel Schr\"odinger equation
with smooth phenomenological potentials. This lack of
exactly-solvable potentials can be related to the poor knowledge
of the scattering inverse problem (i.e., the construction of a
potential in terms of its bound- or scattering-state physical
properties) in the coupled-channel case with threshold
differences~\cite{chadan:89}. In Ref.~\cite{cox:64} however, an
exactly-solvable coupled-channel potential with threshold
differences is derived, two remarkable features of which are the
compact expressions provided both for the potential and for its
Jost matrix. Since the Jost-matrix completely defines the bound-
and scattering-state properties of a potential
model~\cite{newton:82,vidal:92a}, such an analytical expression
seems very promising in the context of the scattering inverse
problem.

The work of Cox has however received little attention, probably because it is plagued by two problems.
First, the way of getting the potential is rather complicated and mysterious:
the paper mostly consists in a check that the provided analytical expression
for the solutions satisfies the coupled-channel Schr\"odinger equation with
the provided analytical expression for the potential.
Not much information is given on how these expressions were obtained,
which makes any generalization of the method impossible.
The second problem, already stressed in Ref.\
\cite{cox:64}, is that, despite the compact expression of the Jost matrix,
calculating the corresponding bound- and resonant-state properties is a difficult
task because these states correspond to zeros of the {\em determinant}
of the Jost matrix in the intricate structure of the energy Riemann sheet, which has a multiplicity $2^N$ for $N$ channels.

The first problem was solved recently, when it was realized that the
Cox potential, at least in its simplest form ($q=1$ in Ref.~\cite{cox:64}),
can be obtained by a single supersymmetric transformation of the zero potential
\cite{sparenberg:06,samsonov:07}.
This leads to a much simpler derivation of this potential and naturally
enables several generalizations of it; in particular, the initial potential
is now arbitrary.
The transformation used to get this result belongs to a category
of supersymmetric transformations not much used up to now, namely
transformations that do not respect the boundary condition at the
origin (the so-called {\em non-conservative} transformations, see e.g.\ Ref.~\cite{sparenberg:06}):
a solution of the initial potential vanishing at the origin is
transformed into a solution of the transformed potential which is finite at the origin.
This feature makes the transformation of the Jost matrix more complicated
to calculate than for usual conservative transformations but it is also
the key to get potentials with non trivial coupling.
In Sec.~\ref{sec:Cox} below, we give several alternative expressions for
the Cox potential and explicitly make the link between the supersymmetric
derivation and the expressions found in Ref.~\cite{cox:64}, for an arbitrary
number of channels $N$.
With respect to Refs.~\cite{sparenberg:06,samsonov:07}, this result is new
as these references mostly concentrate on generalizations of the Cox potential
allowed by supersymmetric quantum mechanics and on $N=2$ examples.
We also show in Sec.~\ref{sec:Cox} that the Cox potential contains
the maximal number of arbitrary parameters allowed by a single
non-conservative supersymmetric transformation,
which makes it the most interesting potential from the point
of view of the scattering inverse problem;
we also derive a new necessary and sufficient condition for the regularity of the potential.

The second problem, i.e., the calculation of bound- and scattering-state
properties from the analytical Jost function, is touched upon
in Sec.~\ref{3}.
There, the discussion is limited to $N=2$ (a case complicated enough from the mathematical point of view but very rich already from the physical point of view).
First, the number of bound states and resonant states is studied geometrically in terms of the potential parameters, as well as the necessary and sufficient condition for a regular potential (several mistakes made in Ref.~\cite{cox:64}, in particular regarding the number
of bound states, are corrected in passing).
Second, schematic inverse problems are solved, where the potential parameters are expressed in terms of physical quantities like bound-state energies, resonance energy and width;
third, the low-energy behavior of the open-channel scattering matrix is studied, with ultracold gases in mind.
This discussion makes possible a first practical use of this potential as a schematic model of atom-atom magnetic Feshbach resonances, which is described in Sec.~\ref{4}.
There, an exactly-solvable model is established in cases where a magnetic Feshbach resonance interplays with a potential resonance, which results in a large background scattering length, either positive (interplay with a bound state) or negative (interplay with a virtual state). This physical context is mostly inspired by Ref.~\cite{marcelis:04}.
Sec.~\ref{sec:conclusion} finally summarizes our findings and discusses
possible extensions of them, in particular to other fields of physics where coupled-channel models are known to play an important role.

\section{\label{sec:Cox} The Cox potential from supersymmetric quantum mechanics}

Let us first summarize the notations used below for coupled-channel scattering theory
\cite{taylor:72,newton:82,vidal:92a}.
We consider a multichannel radial Schr\"odinger equation
that reads in reduced units
\begin{equation}\label{schr}
H\psi(k,r)=K^2\psi(k,r),
\end{equation}
with
\begin{equation}
H=-\frac{d^2}{d r^2}+V,
\end{equation}
where $r$ is the radial coordinate, $V$ is an $N\times N$ real symmetric matrix,
and $\psi$ may be either a matrix-valued or a vector-valued solution.
By $k$ we denote a point in the space ${\mathbb C}^N$,
$k=\left\{k_1,\ldots,k_N\right\}$,
$k_i\in \mathbb C$.
A diagonal matrix with non-vanishing entries $k_i$ is written as
$K=\mbox{diag}(k)=\mbox{diag}(k_1,\ldots,k_N)$.
The complex wave numbers $k_i$ are related to the center-of-mass energy $E$
and the channel thresholds $\Delta_1,\dots, \Delta_N$,
which are supposed to be different from each other,
by
\begin{equation}\label{thrE}
k_i^2=E-\Delta_i\,.
\end{equation}
For simplicity, we assume here that the different channels have
equal reduced masses, a case to which the general situation can
always be formally reduced \cite{newton:82}.
We also assume potential $V$ to be short-ranged at infinity and to support a finite number $M$ of bound states.
Under such assumptions,
the Schr\"odinger equation has two $N\times N$ matrix-valued Jost solutions
which allow one to construct the Jost matrix $F(k)$
defining both scattering and bound-state properties.
The scattering matrix, which is symmetric, reads
\begin{eqnarray}
S(k) & = & K^{-1/2}F(-k)F^{-1}(k)K^{1/2} \nonumber \\
& = & K^{1/2}[F^{-1}(k)]^T F^T(-k)K^{-1/2},
\label{S}
\end{eqnarray}
with $T$ meaning transposition and $-k=\{-k_1,\dots,-k_N\}$.
The zeros of the determinant of the Jost matrix, which are defined by $\det F(k)\equiv 0$, thus correspond to poles of all the elements of the scattering matrix.
Bound states correspond to such zeros $k_m$, with $m=1,\dots,M$, lying on the positive imaginary $k_i$ axes for all channels: $k_{mi}= i \kappa_{mi}$ with $\kappa_{mi} \ge 0$ and $i=1,\dots,N$.
The corresponding energies, $E_m=-\kappa_{mi}^2+\Delta_i$, lie below all thresholds.
For simplicity, we call virtual state any other zero of the Jost-matrix determinant corresponding to a real energy below all thresholds, but not lying on all the positive imaginary $k_i$ axes.
Finally, we call resonance any zero of the Jost-matrix determinant not lying on the imaginary $k_i$ axes, hence corresponding either to a complex energy or to a real energy above at least one threshold.
Note that for a resonance to have a visible impact on the physical scattering matrix it should be located sufficiently close to the real axis.

Let us then summarize the main results from supersymmetric quantum mechanics
in the coupled-channel case~\cite{amado:88a,amado:88b}.
Starting from an initial potential $V$ and its solutions $\psi$,
a supersymmetric transformation allows the construction of a new potential
\begin{equation}\label{Vt}
\tilde{V}(r)=V(r)-2 U'(r)
\end{equation}
with solutions
\begin{equation}\label{psit}
\tilde{\psi}(k,r)=\left[-\frac{d}{dr}+U(r)\right]\psi(k,r)\,,
\end{equation}
where the so-called {\rm superpotential} $U$ is expressed in terms of a
square matrix $\sigma$ by
\begin{equation}\label{U}
U(r)=\sigma'(r) \sigma^{-1}(r)\,.
\end{equation}
Matrix $\sigma$ is called the factorization solution; it is a solution of the initial
Schr\"odinger equation
\begin{equation}
H \sigma(r) = -{\cal K}^2 \sigma(r)\,,
\end{equation}
where ${\cal
K}=\mbox{diag}(\kappa)=\mbox{diag}(\kappa_1,\dots,\kappa_N)$ is a
diagonal matrix called the factorization wave number, which
corresponds to an energy $\cal E$ lying below all thresholds,
called the factorization energy. The entries of $\cal K$ thus
satisfy ${\cal E}=-\kappa_i^2+\Delta_i$; by convention, we choose
them positive: $\kappa_i>0$. Equation~\eqref{psit} implies that
all the physical properties of the transformed potential can be
expressed in terms of those of the initial potential, in
particular its Jost matrix and scattering matrix.

Let us now apply these results to a vanishing initial potential $V=0$,
for which the Jost matrix and scattering matrix are identity, $S(k)=F(k)=I$.
For a given factorization energy,
the most general real symmetric superpotential depends on
an $N$-dimensional real symmetric matrix of arbitrary parameters,
i.e., on $N(N+1)/2$ real arbitrary parameters~\cite{samsonov:07}.
When $V=0$, the corresponding factorization solution can be written as
\begin{subequations}
\label{sigCox}
\begin{eqnarray}
\sigma(r)=\cosh(\kappa r) + {\cal K}^{-1} \sinh(\kappa r) U_0
\label{sigU} \\
 =(2{\cal K})^{-1} [ \exp(\kappa r) ({\cal K} + U_0)
 + \exp(-\kappa r) ({\cal K} - U_0)],
\label{sigexp}
\end{eqnarray}
\end{subequations}
which ensures that the resulting potential $\tilde{V}$ is regular at the origin,
and where
the arbitrary parameters explicitly appear as the value of the (symmetric)
superpotential at the origin, $U_0 \equiv U(0)$;
$ \exp(\pm \kappa r)$, $ \cosh(\kappa r)$ and $\sinh(\kappa r)$ are diagonal
matrices with entries
$ \exp(\pm \kappa_i r)$, $ \cosh(\kappa_i r)$ and $\sinh(\kappa_i r)$
respectively.
According to Ref.~\cite{samsonov:07},
when ${\cal K}+U_0$ is invertible,
the transformed Jost matrix reads
\begin{equation}\label{FtCox}
\tilde{F}(k)=({\cal K}-i K)^{-1}(U_0-i K).
\end{equation}
This is the Jost function obtained by other means
in Ref.~\cite{cox:64} in the case $q=1$.
However, it was not realized there that the corresponding potential could be simply
expressed in terms of a solution matrix $\sigma$, using Eqs.~\eqref{Vt} and~\eqref{U}.
In that reference, a compact expression for the potential is found
[see Eq.~\eqref{VCox} below]
but writing~\eqref{sigCox} and~\eqref{U} is much more
elegant because both the potential~\eqref{Vt}
and its Jost function~\eqref{FtCox} are expressed
in terms of the same parameter matrix $U_0$.
Nevertheless, this procedure also presents several disadvantages:
calculating the potential requires several matrix operations
(inversion, product, derivations);
moreover, the parameters in $U_0$ should be chosen so that
 the factorization solution is invertible for all $r$,
 a condition not easily checked on Eqs.~\eqref{sigCox}.

Let us now derive an alternative form for the factorization solution,
which solves both these inconveniences.
In Ref.~\cite{samsonov:07}, the possibility of rank
$({\cal K} +U_0) < N$ in Eq.~\eqref{sigexp} has been studied,
which leads to an interesting asymptotic behavior of the superpotential
but which reduces the number of parameters in the model.
Here, in order to keep the maximal number of arbitrary parameters in the potential,
we choose ${\cal K} +U_0$ invertible.
The factorization solution~\eqref{sigexp} can then be multiplied on the right by
 $2 ({\cal K} +U_0)^{-1} {\cal K}^{1/2}$,
which leads to the factorization solution
\begin{equation}
\sigma(r) = {\cal K}^{-1/2} \left[ \exp(\kappa r) + \exp(-\kappa r) X_0 \right].
\label{sigX0}
\end{equation}
According to Eq.~\eqref{U}, the superpotential, and hence the transformed potential,
is unaffected by this multiplication.
The symmetric matrix $X_0$ now contains all the arbitrary parameters.
The link between the two sets of parameters is given by
\begin{eqnarray}\label{XU}
X_0 & = & {\cal K}^{-1/2}
({\cal K} - U_0) ({\cal K} + U_0)^{-1}{\cal K}^{1/2}\,,
 \\
U_0 & = & {\cal K}^{1/2} (I-X_0) (I+X_0)^{-1}{\cal K}^{1/2}\,.
\label{UX}
\end{eqnarray}

Equation~\eqref{sigX0} can also be written as
\begin{equation}\label{sigX}
\sigma(r) = {\cal K}^{-1/2} \left[ I + X(r) \right] \exp(\kappa r)\,,
\end{equation}
where
\begin{equation}\label{X}
X(r)= \exp(-\kappa r) X_0 \exp(-\kappa r).
\end{equation}
With respect to writing~\eqref{sigU} and~\eqref{sigexp},
Eq.~\eqref{sigX} presents several advantages.
First, it allows for a simple calculation of the superpotential
\begin{eqnarray}
U(r) & = & {\cal K} - 2 {\cal K}^{1/2} X(r) [I+X(r)]^{-1} {\cal K}^{1/2} \nonumber \\
& = & -{\cal K} + 2 {\cal K}^{1/2} [I+X(r)]^{-1} {\cal K}^{1/2}\,. \label{UrX}
\end{eqnarray}
The last expression is particularly convenient since the $r$ dependence is limited
to one factor of the second term; the potential can thus be explicitly written as
\begin{widetext}
\begin{eqnarray}
\tilde{V}(r) & = & 4 {\cal K}^{1/2} [I+X(r)]^{-1} X'(r) [I+X(r)]^{-1} {\cal K}^{1/2} \nonumber \\
& = & - 4 {\cal K}^{1/2} \left(e^{\kappa r} +
X_0 e^{-\kappa r}\right)^{-1} (X_0 {\cal K}+{\cal K} X_0)
\left(e^{\kappa r} + e^{-\kappa r} X_0 \right)^{-1} {\cal K}^{1/2}\,.
\end{eqnarray}
\end{widetext}
The last expression is exactly equivalent to Eq.\ (4.7)
of Ref.~\cite{cox:64} for $q=1$,
which reads
\begin{eqnarray}
\tilde{V}(r)&=& 2 e^{-\kappa r} \left[I-A (2{\cal K})^{-1} e^{-2 \kappa r}\right]^{-1}
(A{\cal K}+{\cal K}A) \nonumber \\
& \times & \left[I-e^{-2 \kappa r} (2{\cal K})^{-1}A \right]^{-1} e^{-\kappa r}\,,
\label{VCox}
\end{eqnarray}
provided one defines matrix $A$ as
\begin{eqnarray}
A & = & -2 {\cal K}^{1/2} X_0 {\cal K}^{1/2} \nonumber \\
& = & - 2 ({\cal K} - U_0)({\cal K}+U_0)^{-1} {\cal K}\,.
\label{AU}
\end{eqnarray}

The second advantage of writing~\eqref{sigX}
 is that it easily leads to a necessary and sufficient condition
 on the parameters to get a potential without singularity
 at finite distances.
 This condition is positive definiteness of matrix
 $I+X_0$:
 \begin{equation}\label{posX}
 I+X_0>0\,.
 \end{equation}

The potential has a singularity when $\sigma(r)$ is noninvertible,
i.e., when $\det[I+X(r)]$ vanishes for some $r$.
Using Eq.~\eqref{X}, we find that this is equivalent to
the existence of $r_0\ge0$ such that
$\det Y(r_0)=0$ with $Y(r)=\exp(2 \kappa r)+X_0$.
Assume now that $\det Y(r)\ne0$ $\forall r\ge0$.
Since $\det Y(r)=\prod_{i=1}^N y_i(r)$ where $y_i(r)$ are
the eigenvalues of $Y(r)$, we conclude that
  $y_i(r)\ne0$ for all $i=1,\ldots,N$ and $r\ge0$.
But since for sufficiently large $r$, $X_0$ becomes a small perturbation
to $\exp(2\kappa r)$, all eigenvalues of $Y(r)$ should
be
positive for $r\ge0$ and in particular at $r=0$, thus proving the
necessary character of the above condition.

The sufficiency follows from the observation that $Y(r)$
is positive definite for any $r\ge0$, together with  $Y(0)=I+X_0$.
Indeed,
if $Y(r)$ is positive definite, the
inequality $\langle q|Y(r)|q\rangle >0$ holds for any $q\in L_N$.
Here $\langle p\,|q\rangle=\sum_{i=1}^Np^*_iq_i $ is the usual
inner product in the $N$-dimensional complex linear space $L_N$, with
$p_i$, $q_i$ being coordinates of the vectors $p,q\in L_N$
with respect to an orthonormal basis.
But since
$\langle q|Y(r)|q\rangle =\langle q|X_0|q\rangle+
\langle q|\exp(2 \kappa r)|q\rangle
\ge \langle q|X_0|q\rangle+
\langle q|q\rangle=\langle q|X_0+I|q\rangle$
[we recall that $r\ge0$, $\kappa_i>0$ and $\exp(\kappa r)$ is a diagonal matrix with entries $\exp(\kappa_i r)$],
positive definiteness of $I+X_0$ implies positive definiteness of
$Y(r)$ for $r\ge0$.

Having established this condition on $X_0$, one can get the condition in terms of $U_0$, using Eq.~\eqref{XU}.
Since
\begin{equation}
 I+X_0= 2 {\cal K}^{1/2} ({\cal K} + U_0)^{-1} {\cal K}^{1/2},
\end{equation}
the necessary and sufficient condition to get a regular potential is positive definiteness of matrix ${\cal K}+U_0$:
\begin{equation}\label{posU}
{\cal K}+U_0>0\,.
\end{equation}
Since the (diagonal) elements of $\cal K$ are positive and increase when the factorization energy decreases,
this condition has a simple interpretation: it just puts some upper limit on the factorization energy.

Finally, Eq.~\eqref{AU} shows that the condition
det $A\ne0$ required in Ref.~\cite{cox:64} is not required here.
In Cox' paper, this condition does not appear in the potential expression,
which is valid in the general case, but only in the derivation of the proof;
the fact that this condition is not required here illustrates the efficiency
of the supersymmetric formalism.
Equation~\eqref{AU} also implies that rank $({\cal K}+U_0) < N$ corresponds
to det $A=\infty$, a case also not considered in Ref.~\cite{cox:64}.
The supersymmetric treatment, on the contrary, allows this case~\cite{sparenberg:06,samsonov:07};
our approach thus subsumes the results of Ref.~\cite{cox:64} in several respects.

\section{General properties of the $2\times2$ Cox potential\label{3}}

Having established a connection between the Cox potential and supersymmetric quantum mechanics, we now proceed in a more detailed analysis of its properties
for the simplest particular case $N=2$.
As is happens, this case is not only complicated enough to deserve a dedicated analysis, but also sophisticated enough to make the solution of several interesting inverse problems possible.

\subsection{Explicit expression of the potential}
%
For $N=2$, the arbitrary parameters entering the Cox potential are the entries of the superpotential matrix at the origin,
\be\label{sp0}
U_0 \equiv U(0)=  \left(
\begin{array}{cc}
\a_1 & \b \\[.5em] \b & \a_2
\end{array}
\right), \ee
and the factorization energy $\cal E$.
The corresponding factorization wave number, $\kappa=(\kappa_1, \kappa_2)$, is made of two positive parameters $\kappa_1$ and $\kappa_2$ which are not independent of each other: they should satisfy the ``threshold condition'' [see Eq.~\eqref{thrE}]
\begin{equation}
 \kappa_2^2-\kappa_1^2=\Delta.
 \label{threshold-}
\end{equation}
Here and in what follows we put for convenience $\Delta_1=0$, $\Delta_2=\Delta>0$.

In terms of these parameters, the necessary and sufficient condition for a regular potential, i.e., ${\cal K}+U_0$ positive definite, can be written for instance
\begin{subequations}
\label{nsc12}
\begin{eqnarray}
 \kappa_1 & > & -\alpha_1, \label{nsc1} \\
 \kappa_2 & > & \frac{\b^2}{\kappa_1+\a_1}-\a_2.
\label{nsc2}
\end{eqnarray}
\end{subequations}
This puts an upper limit on the factorization energy in terms of the parameters appearing in $U_0$ [see Eq.~\eqref{nonsp1} and Fig.~\ref{figS} below].

Two explicit expressions for the superpotential are given in Ref.~\cite{samsonov:07}.
Using Eqs.~\eqref{Vt} and~\eqref{UrX}, one gets what is probably the simplest possible explicit expression for the potential itself:
\begin{widetext}
\begin{subequations}
\label{vtcox}
\begin{eqnarray}
\tilde{v}_{11} & = & -8 \kappa_1 e^{-2 \kappa_1 r} \
\frac{x_{11} \kappa_1 +\left[2 x_{11} x_{22} \kappa_1 - x_{12}^2 \left(\kappa_1+\kappa_2\right)\right] e^{-2 \kappa_2 r}
+ x_{22} \left(x_{11} x_{22} - x_{12}^2\right) \kappa_1 e^{-4 \kappa_2 r}}
{\left[1+x_{11} e^{-2\kappa_1 r} +x_{22} e^{-2\kappa_2 r} + \left(x_{11} x_{22} - x_{12}^2\right) e^{-2(\kappa_1+\kappa_2) r} \right]^2}, \label{vt11} \\
\tilde{v}_{12} & = & -4 x_{12} \sqrt{\kappa_1 \kappa_2} e^{-(\kappa_1+\kappa_2) r} \times \nonumber \\
 && \frac{\kappa_1+\kappa_2+x_{11} (\kappa_2-\kappa_1) e^{-2\kappa_1 r} +x_{22} (\kappa_1-\kappa_2) e^{-2\kappa_2 r} - \left(x_{11} x_{22} - x_{12}^2\right) (\kappa_1+\kappa_2) e^{-2(\kappa_1+\kappa_2) r}}
{\left[1+x_{11} e^{-2\kappa_1 r} +x_{22} e^{-2\kappa_2 r} + \left(x_{11} x_{22} - x_{12}^2\right) e^{-2(\kappa_1+\kappa_2) r} \right]^2}.
\end{eqnarray}
\end{subequations}
\end{widetext} The element $\tilde{v}_{22}$ is obtained from
Eq.~\eqref{vt11} by the replacement $\kappa_1 \leftrightarrow
\kappa_2$ and $x_{11} \leftrightarrow x_{22}$. Here, we have used
the symmetric matrix
\begin{equation}
X_0 =  \left(
\begin{array}{cc}
x_{11} & x_{12} \\[.5em] x_{12} & x_{22}
\end{array}
\right),
\end{equation}
which is related to matrix~\eqref{sp0} by Eqs.~\eqref{XU} and~\eqref{UX}.
In the following, as we are mostly interested in the Jost-matrix properties, we shall rather use matrix $U_0$.

\subsection{\label{subsec:zeros} Zeros of the Jost-matrix determinant}

Let us denote for convenience the channel wave numbers as $k_1=k$ and $k_2=p$,
with the threshold condition
\be
k^2-p^2=\Delta.
\label{threshold}
\ee
Then,
according to Eq.~\eqref{FtCox},
the Jost matrix for the Cox potential reads
(see also Refs.~\cite{cox:64,sparenberg:06,samsonov:07})
\be\label{F}
\tilde{F}(k,p)= \left(
\begin{array}{cc}
\frac{k+i\a_1}{k+i\kappa_1} & \frac{i\b}{k+i\kappa_1} \\[.5em] \frac{i\b}{p+i\kappa_2} &
\frac{p+i\a_2}{p+i\kappa_2}
\end{array}
\right).
\ee
The determinant of the Jost matrix coincides with the Fredholm determinant of the
corresponding integral equation~\cite{newton:82}; it reads here

\begin{widetext}
\begin{figure}
\begin{center}
\begin{minipage}{17cm}
\epsfig{file=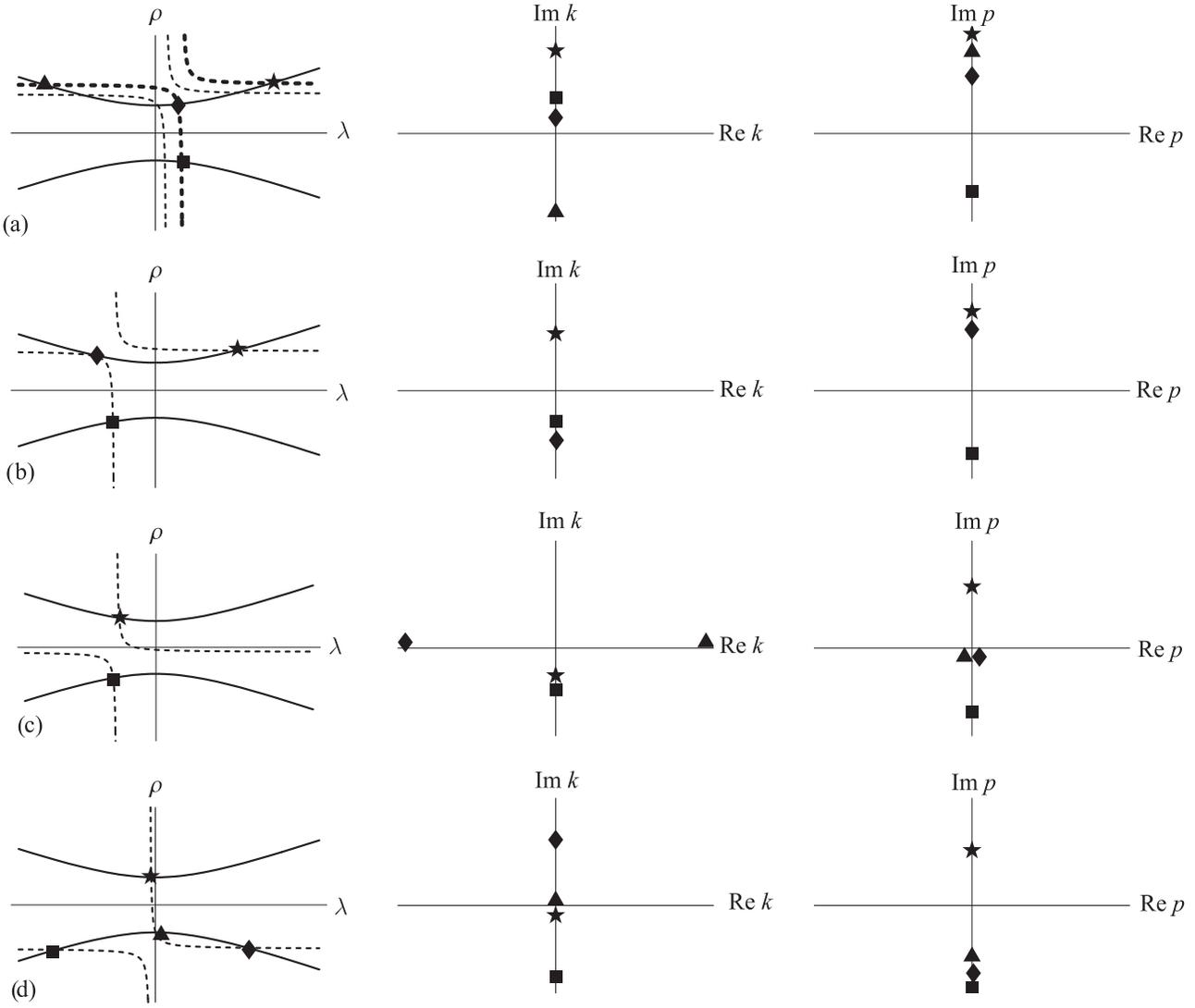, width=17cm}
\caption{\small
Geometrical representation of Eqs.~\eqref{sys1a} (first column,
solid lines) and~\eqref{sys1b} (first column, dashed lines), and
positions of the corresponding roots of system~\eqref{sys} in the
complex $k$ (second column) and $p$ (third column) planes. Various
values of the parameters $\alpha_1$, $\alpha_2$ are chosen, which
imply various numbers of bound, virtual and resonant states:
(a)~$\a_1<0$, $\a_2<-\sqrt{\Delta}$, two bound states (star and
diamond), two virtual states (square and triangle), no resonance;
(b)~$\a_1>0$, $\a_2<-\sqrt{\Delta}$, one bound state (star), one
virtual state (square), appearance of a resonance (diamond);
(c)~$\a_1>0$, $\a_2>0$, no bound state, two virtual states (star
and square), one resonance (triangle and diamond); (d)~$\a_1>0$,
$\a_2>\sqrt{\Delta}$, no bound state, no resonance, four virtual
states. Increase of either $\a_1$ or $\a_2$ leads to: (a)~(thin
dashed lines) disappearance of a bound state; (b)~appearance of
the resonance. \label{fig1}}
\end{minipage}
\end{center}
\end{figure}
\end{widetext}

\be\label{det1}
f(k,p)\equiv\mbox{det}\tilde{F}(k,p)=
\frac{(k+i\a_1)(p+ia_2)+\b^2}{({k+i\kappa_1})(p+i\kappa_2)}.
\ee

The zeros of this Jost determinant in the $k$ and $p$ complex planes, which correspond to bound, virtual or resonant states, are functions of the parameters $\a_1$, $\a_2$ and $\b$ only.
From here and the threshold condition~\eqref{threshold},
follows the system of equations for finding the zeros of $f(k,p)$,
\begin{subequations}
\label{sys}
\begin{eqnarray}
&& k^2-p^2=\Delta, \\
&& (k+i\a_1)(p+i\a_2)+\b^2=0, \label{sys1}
\end{eqnarray}
\end{subequations}
which is equivalent to the fourth-order algebraic equation:
\be\label{k4}
k^4+ia_1k^3+a_2k^2+ia_3k+a_4=0\,,
\ee
where
\begin{subequations}
\begin{eqnarray}
a_1 & = & 2\a_1, \\
a_2 & = & \a_2^2-\a_1^2-\Delta, \\
a_3 & = & 2[\a_1(\a_2^2-\Delta)-\a_2\b^2], \\
a_4 & = & -\a_1^2(\a_2^2-\Delta)+2\a_2\b^2\a_1-\b^4.
\end{eqnarray}
\end{subequations}
We notice that after substitution $k=i\lambda$, Eq.~\eqref{k4}
becomes an algebraic equation in $\lambda$ with real coefficients.

Its four roots are thus either real numbers, which correspond to real negative energies (bound or virtual states),
or mutually-conjugated complex numbers, which correspond to mutually-conjugated complex energies (resonant states).
Basing on this property, we will use in what follows a geometric
representation of the system of equations which allows for a
visualization of the zeros
of $f(k,p)$
in the parameter space.

Let us first consider bound and virtual states, which correspond to solutions of system~\eqref{sys} with $k$ and $p$ purely imaginary.
After substitution $k=i\lambda$, $p=i\rho$, with $\lambda$ and $\rho$ real,
these equations define two hyperbolas in the $(\lambda,\rho)$-plane,
\begin{subequations}
\label{syslr}
\begin{eqnarray}
\label{sys1a}
&& \rho^2-\lambda^2 = \Delta\,,\\
&&  (\lambda+\a_1)(\rho+\a_2) = \b^2,
\label{sys1b}
\end{eqnarray}
\end{subequations}
the positions of which are defined by the values of the parameters
$\a_1$, $\a_2$, $\b$ and $\Delta$.

The roots of system~\eqref{syslr} that correspond to bound and virtual states are the intersection points of these hyperbolas.
Different possibilities of hyperbola locations are shown in Fig.~\ref{fig1}.
The solid-line hyperbola corresponds to the threshold condition~\eqref{sys1a};
its semi-major axis is $\sqrt{\Delta}$ and its slant asymptotes are given by $\rho=\pm \lambda$.
The dashed-line hyperbola corresponds to Eq.~\eqref{sys1b};
its asymptotes are given by $\lambda=-\alpha_1$ and $\rho=-\alpha_2$.
The abscissa (resp., ordinate) of a crossing point in the $(\lambda,\rho)$-plane gives
the position of the corresponding zero on the imaginary axis in the $k$-plane (resp., $p$-plane),
as shown in the second (resp., third) column of Fig.~\ref{fig1}.
Bound states correspond to $\lambda,\rho>0$,
i.e., to intersection points laying in the first quadrant of the $(\lambda,\rho)$-plane,
while virtual states correspond to intersections in the second, third and fourth quadrants.
It is clearly seen on Fig.~\ref{fig1} that the two hyperbolas~\eqref{sys1a} and~\eqref{sys1b}
cross in either two or four points.
Moreover, they can have zero, one or two intersections in the first quadrant,
which means that the potential has either zero, one or two bound states.
This contradicts Ref.~\cite{cox:64}, where it is said that the potential does never support bound states.
Since Eq.~\eqref{k4} is fourth order, when the hyperbolas cross in four points,
the Jost determinant does not have any other zero; on the other hand,
when the hyperbolas cross in only two points, the Jost determinant has two other zeros,
which have to form a mutually-conjugated complex pair, as seen above.
This last case corresponds to a resonance, as illustrated by Fig.~\ref{fig1}(c),
where the hyperbolas only have two intersection points in the $(\lambda,\rho)$-plane
and a pair of complex roots appears in the complex $k$ and $p$ planes.
The potential thus has either zero or one resonance.
The intermediate case of three intersection points for the hyperbolas [Fig.~\ref{fig1}(b)]
corresponds to the presence of a multiple root of Eq.~\eqref{k4}, which
lies in an unphysical sheet
(${\rm Im} k<0,\,{\rm Im}p>0$ or ${\rm Im} k>0,\,{\rm Im}p<0$)
of the Riemann energy surface;
this case corresponds to a transition between a one-resonance and a two-virtual-state situation.

One sees that the parameters $\a_1$ and $\a_2$ determine the position of hyperbola~\eqref{sys1b} and,
hence, the number of bound states $n_b$ (0, 1 or 2) and of resonances $n_r$ (0 or 1).
Let us now determine, for fixed values of $\b$ and $\Delta$,
the domains in the plane of parameters $\mathbb{A}=(\a_1,\a_2)$ with constant values of $n_b$ and $n_r$.
To find domains in $\mathbb{A}$ where system~\eqref{syslr}
has two complex conjugated roots (one
resonance), we consider the case where the hyperbolas have a common tangent point,
as illustrated by Fig.~\ref{fig1}(b).
One can see that the decrease of either $\a_1$ or $\a_2$ leads to the disappearance of the resonance,
while the increase of either $\a_1$ or $\a_2$ leads to the appearance of the resonance.
We define the parametric curves $[\a_1(\lambda_0,\rho_0),\alpha_2(\lambda_0,\rho_0)]$
in plane $\mathbb{A}$ by shifting the
tangent point $(\lambda_0,\rho_0)$ along the hyperbola $\rho^2-\lambda^2=\Delta$.
These curves limit domains in $\mathbb{A}$ with either zero or two complex roots.
To find them, we use the two conditions corresponding to the common tangent
point $(\lambda_0,\rho_0)$
\begin{subequations}
\label{sysres}
\begin{eqnarray}
&& \rho_0=\frac{\b^2}{\lambda_0+\a_1}-\a_2=
\pm\sqrt{\lambda_0^2+\Delta}, \\
 && \left.
\frac{d\rho}{d\lambda}\right|_{\lambda=\lambda_0}=
-\frac{\b^2}{(\lambda_0+\a_1)^2}
=\pm\frac{\lambda_0}{\sqrt{\lambda_0^2+\Delta}}\,.
\end{eqnarray}
\end{subequations}
The upper signs correspond to $\lambda_0<0$ (tangent point in the second quadrant) while the lower signs correspond to $\lambda_0>0$ (tangent point in the fourth quadrant).
We can solve system~\eqref{sysres} with respect to $\a_1$ and $\a_2$:
\begin{subequations}
\label{rc1}
\begin{eqnarray}
\!\!\a_1(\lambda_0)&=&
\pm\frac{\b}{\sqrt{|\lambda_0}|}(\lambda_0^2+\Delta)^{1/4}-\lambda_0\,,\\
\!\!\a_2(\lambda_0)&=&
\pm\frac{\b\sqrt{|\lambda_0|}}{(\lambda_0^2+\Delta)^{1/4}}+{\rm sign}(\lambda_0)\sqrt{\lambda_0^2
+\Delta}
\,.
\end{eqnarray}
\end{subequations}

\begin{figure}
\begin{center}
\begin{minipage}{8.6cm}
\epsfig{file=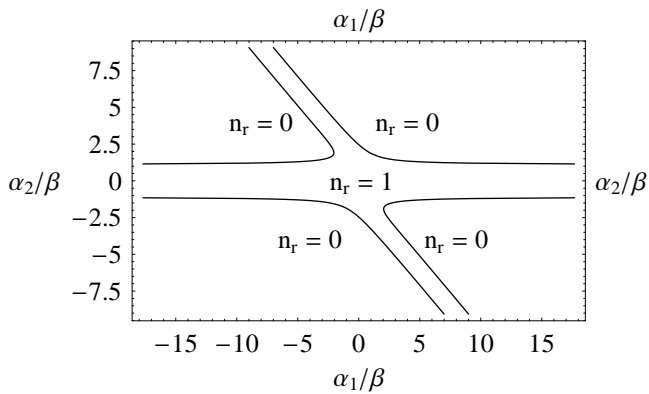, width=8.6cm}
\caption{\small
Parametric curves in terms of dimensionless parameters defined by Eqs.~\eqref{rc2}
in plane $\widetilde{\mathbb{A}}=(\widetilde{\alpha}_1,\widetilde{\alpha}_2)$
for $\Delta/\b^2=1.2$.
The left-hand-side curves correspond to the lower signs in the equations,
while the right-hand-side curves correspond to the upper signs.
The number of resonances, $n_r$, is indicated in each domain of the plane. \label{fig2r} }
\end{minipage}
\end{center}
\end{figure}

It should be noted that the Schr\"odinger equation with the Cox
potential has the following scale invariance:
\begin{subequations}
\begin{eqnarray}
\a_{1,2} &\to& \gamma\a_{1,2}\,,\qquad \Delta \to \gamma^2\Delta\,,\\
\kappa_{1,2} &\to& \gamma \kappa_{1,2}\,,\qquad \b \to \gamma \b\,, \\
r &\to& r/\gamma\,,
\end{eqnarray}
\end{subequations}
which leaves $\Delta_d=\Delta/\b^2$ invariant. Hence, we may put $\Delta=1$ without losing generality. This
choice is equivalent to measuring energies in units of $\Delta$. It is convenient to express
equations \eqref{rc1} in terms of dimensionless variables $\a_i/\b$, $\Delta_d=\Delta/\b$, $\lambda_0\rightarrow\lambda_0/\b$:
\begin{subequations}
\label{rc2}
\begin{eqnarray}
\!\!\!\!\!\!\!\! \frac{\a_1}{\b}(\lambda_0)&=&
\pm\frac{1}{\sqrt{|\lambda_0|}}(\lambda_0^2+\Delta_d)^{1/4}-\lambda_0\,,\\
\!\!\!\!\!\!\!\! \frac{\a_2}{\b}(\lambda_0)&=&
\pm\frac{\sqrt{|\lambda_0|}}{(\lambda_0^2+\Delta_d)^{1/4}}+{\rm sign}(\lambda_0)\sqrt{\lambda_0^2
+\Delta_d}.
\end{eqnarray}
\end{subequations}
These four solutions [taking into account ${\rm sign}(\lambda_0)$] can be considered as four parametric curves in plane
${\mathbb A}=(\a_1/\b,\a_2/\b)$,
which separate the plane in five regions (one inner region and four outer regions, see Fig.~\ref{fig2r}).

In the inner region,
the Jost determinant has two complex roots $k_{1,2}=\pm k_r+ik_i$
and, hence, these values of parameters $\a_1,\a_2$ correspond to
one resonance ($n_r=1$).
In the four outer regions, the Jost determinant has purely-imaginary roots, hence $n_r=0$.
The curves in Fig.~\ref{fig2r} tend asymptotically to straight
lines which are defined as the limits for
$\lambda_0\rightarrow 0$ and $\lambda_0\rightarrow\pm\infty$.
As a result, one finds for all branches
two horizontal asymptotes $\a_2/\b=\pm\sqrt{\Delta_d}$
and three slant asymptotes defined by $\a_2/\b=-\a_1/\b$
(for the curves in the second and fourth quadrants) and  $\a_2/\b=-\a_1/\b\pm 2$
(for the curves in the first and third quadrants, respectively).

\begin{figure}
\begin{center}
\begin{minipage}{8.6cm}
\epsfig{file=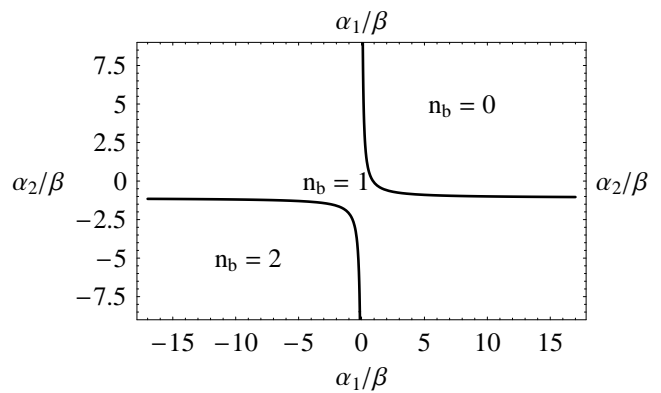, width=8.6cm}
\caption{\small Curves~\eqref{bsz} in plane $\widetilde{\mathbb{A}}$ in terms of
dimensionless parameters for
$\Delta/\b^2=1.2$.
The number of bound states, $n_b$,
is indicated in each domain of the plane.\label{fig2b} }
\end{minipage}
\end{center}
\end{figure}

Consider now the case where the hyperbolas cross at the point
$\lambda_0=0$, $\rho_0=\sqrt{\Delta}$
[see the thin dashed lines in Fig.~\ref{fig1}(a)].
After a small decrease of either $\a_1$ or $\a_2$,
the number of positive roots, i.e., of bound states, increases by one unit.
Hence, assuming $\lambda_0=0$ and $\rho_0=\sqrt{\Delta}$ in system~\eqref{syslr},
we get the curves
\be\label{bsz}
\a_1\left(\a_2+\sqrt{\Delta}\right)-\b^2=0,
\ee
which define three domains in the plane of parameters $\mathbb A$,
where Eqs.~\eqref{syslr} have different number of positive roots
(see Fig.~\ref{fig2b}).
One can directly check that the number $n_b$ of bound states may be calculated as a function of the parameters as
\be\label{nb}
 n_b=1+\frac{1}{2}\left(I_1-1\right)I_2,
 \ee
where the quantities
\begin{subequations}
\label{ind12}
\begin{eqnarray}
\label{ind1}
I_1 &=&{\rm sign}
\left(\b^2-\a_1\sqrt{\Delta}-\a_1\a_2\right)\cdot1, \\
\label{ind2}
I_2 &=&{\rm sign}(\a_2+\sqrt{\Delta})\cdot1
\end{eqnarray}
\end{subequations}
may be considered as invariants.
For $n_b=0$, one has $I_1=-1$ and $I_2=1$; for $n_b=1$, one has $I_1=1$ and $I_2=\pm 1$; for $n_b=2$, one has $I_1=I_2=-1$.

\begin{figure}
\begin{center}
\begin{minipage}{8.6cm}
\epsfig{file=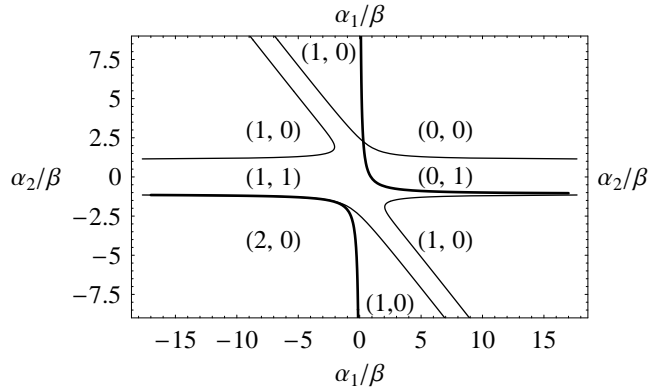, width=8.6cm}
\caption{\small Regions of the ${\mathbb A}$-plane with different numbers of bound states
and resonances, $(n_b,n_r)$, for the Cox potential with $\Delta/\b^2=1.2$.
\label{fig2}}
\end{minipage}
\end{center}
\end{figure}

Let us now summarize our findings on the number of bound states and resonances of the $2\times 2$ Cox potential, by combining Figs.~\ref{fig2r} and~\ref{fig2b} in Fig.~\ref{fig2},
where both $n_b$ and $n_r$ are given for all the possible regions of plane $\mathbb A$.
The border lines of these regions, as already discussed,
correspond to the parametric curves defined by Eqs.~\eqref{rc1},
\eqref{rc2}, and to the curves given by Eq.~\eqref{bsz}. From the
asymptotic behavior of these curves, it is easy to see the global
structure of the zones. For instance, for the case of two bound
states, the hyperbolas in Fig.~\ref{fig1} have to have four
intersection points, which implies that no resonance is present.
This is the reason why the boundary lines between the zones of
bound and resonant states do not cross in the lower-half
$\mathbb{A}$-plane. Moreover, one can see that the topological
structure of these zones does not depend on a particular choice of
a parameter $\Delta_d=\Delta/\b^2$. A change of this parameter only
leads to a deformation of zones, namely, the distance between horizontal asymptotes changes, but does not make any new
intersection point or new boundary line appear.

The case of $\b=0$, $\Delta_d=\infty$ corresponds to uncoupled channels. In this case there are no resonances. Only bound or
virtual states located in different channels may appear (see Sec.~\ref{4}. A.)

\begin{figure}
\begin{center}
\epsfig{file=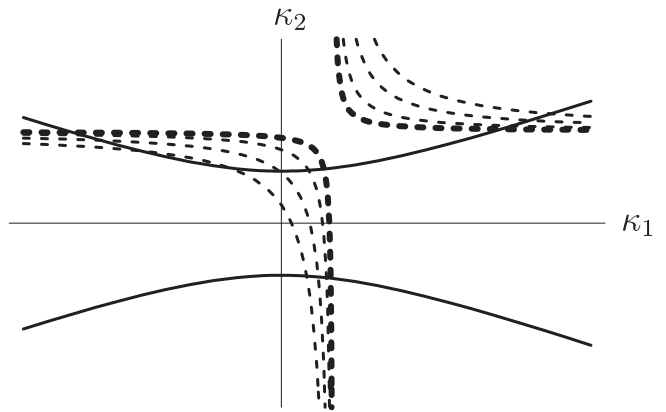, width=8.6cm} \caption{\small Graphical
representation of Eqs.~\eqref{threshold-} (solid lines) and~\eqref{nonsp1} (bold dashed lines for $C=0$, thin dashed lines for $C>0$).
The permitted values of $\kappa_{1,2}$ correspond to the upper-right intersections between the solid and thin-dashed hyperbolas.
After the substitution $\kappa_1 \rightarrow \lambda$, $\kappa_2 \rightarrow \rho$, the solid lines also represent Eq.~\eqref{sys1a} and the bold dashed lines also represent Eq.~\eqref{sys1b}, the first-quadrant intersections of which correspond to bound states.
\label{figS}}
\end{center}
\end{figure}

Up to now, we have excluded the factorization
energy from our analysis because Eqs.~\eqref{syslr} are independent of $\kappa_{1,2}$.
We will now give a geometrical analysis of conditions~\eqref{nsc12},
that have to be imposed on parameters $\kappa_{1}$ and
$\kappa_{2}$ to warranty a regular character of the Cox potential $\forall r\ge0$.
Let us notice that condition~\eqref{nsc2},
\be\label{nonsp} (\kappa_1+\a_1)(\kappa_2+\a_2)-\b^2>0,
\ee is nothing but Eq.~\eqref{sys1b} for $\lambda=\kappa_1$, $\rho=\kappa_2$,
and a shifted value of parameter $\b^2$.
Actually, since the left-hand side of Eq.~\eqref{nonsp} should be positive,
there exists a positive number $C$ (depending on the set of parameters)
such that
\be\label{nonsp1}
(\kappa_1+\a_1)(\kappa_2+\a_2)-(\b^2+C)=0\,,\qquad C>0\,.
\ee
This equation represents a hyperbola with the same asymptotes as hyperbola~\eqref{sys1b} but with a larger distance between its branches.
We thus conclude that the permitted values of $\kappa_{1,2}$ are determined by the intersection points of hyperbolas~\eqref{threshold-} and~\eqref{nonsp1} for arbitrary values of $C$, with the additional condition~\eqref{nsc1}: $\kappa_1>-\alpha_1$.
Figure~\ref{figS} helps to visualize these conditions: the allowed values of the factorization energy correspond to the values of $\kappa_{1,2}$ given by the upper-right intersections between the solid and the thin-dashed hyperbolas.
In Fig.~\ref{figS}, the solid and bold-dashed hyperbolas can also be considered as representing Eqs.~\eqref{syslr}; the bound-state energies then correspond to their (0, 1 or 2) intersections in the first quadrant.
The allowed values of $\kappa_{1,2}$ should thus be larger than the largest values of $\lambda$, $\rho$ corresponding to a bound state.
The necessary and sufficient condition for a regular potential can thus be simply stated as: the factorization energy should be negative and lower than the lowest bound-state
energy, if any.

\subsection{Inversion of the zeros of the Jost-matrix determinant}

To solve a realistic two-channel scattering inverse problem,
it is necessary to express the Cox potential in terms of physical data such as the threshold energy, bound-state energies, resonance energy and width, or scattering data.
While the threshold energy explicitly appears in the expression of the Cox potential as parameter $\Delta$,
the other data are directly related to the positions of the zeros of the Jost-matrix determinant, as seen above.
Ideally, one would thus like to directly express parameters $\a_{1}$, $\a_{2}$, $\b$, and
$\cal E$, which define the Cox potential, in terms of the roots of Eq.~\eqref{k4}.
Certainly, there exist general formulas for the roots of the fourth-order
algebraic equation~\eqref{k4},
but they are very involved and cannot help much in realizing the above program.
Therefore, we propose here an intermediate approach.
Two of the roots of Eq.~\eqref{k4} happen to be easily expressed in terms of parameters $\a_1$ and $\a_2$.
Once two roots are fixed, Eq.~\eqref{k4} reduces to a second-order
algebraic equation for the two other roots, thus providing an implicit but
rather simple mapping between the roots of Eq.~\eqref{k4} and the set of
parameters.

Let us denote by $(k_1,p_1)$ and $(k_2,p_2)$ two zeros of $f(k,p)$.
This imposes some restrictions on the parameters $\a_1$ and $\a_2$.
They are not independent anymore from the other parameters but should be found
as functions of $k_{1,2}$, $p_{1,2}$, and $\b$ from the system of
two equations obtained from Eq.~\eqref{sys1}, written
for $k=k_1$, $p=p_1$ and for $k=k_2$, $p=p_2$.
This system reads
\begin{subequations}
\label{sys11a}
\begin{align}
& (k_1+i\a_1)(p_1+i\a_2)+\b^2=0\,,\\
& (k_2+i\a_1)(p_2+i\a_2)+\b^2=0\,.
\end{align}
\end{subequations}
After eliminating parameter $\a_2$ from system~\eqref{sys11a},
one finds a second-order algebraic equation for $\a_1$,
\be
\a_1^2-\a_1i(k_1+k_2)-k_1k_2+\b^2\frac{\Delta_k}{\Delta_p}=0\,,
\ee
with
\begin{subequations}
\begin{eqnarray}
\Delta_k &=& k_2-k_1,\\
\Delta_p &=& p_2-p_1,
\end{eqnarray}
\end{subequations}
from which follows the two possible choices
\begin{subequations}
\label{alp}
\begin{eqnarray}\label{alp1}
\a_1 &=& \frac{1}{2}\left[i(k_1+k_2)\pm\sqrt{-\Delta_k^2-4\b^2\Delta_k/\Delta_p}\right],\\
\label{alp2}
\a_2 &=& \frac{1}{2}\left[i(p_1+p_2)\mp\sqrt{-\Delta_p^2-4\b^2\Delta_p/\Delta_k}\right].
\end{eqnarray}
\end{subequations}
The upper (resp., lower) sign in Eq.~\eqref{alp2} corresponds to the upper (resp., lower)
sign in Eq.~\eqref{alp1}.
The values of $k_{1,2}$ and $p_{1,2}$ should be chosen so as to
warranty the reality of parameters $\a_{1,2}$.

Now, two roots $k_{1}$ and $k_{2}$ of the fourth-order algebraic equation~\eqref{k4} are fixed.
Therefore, the two remaining roots $k_{3}$ and $k_{4}$
are solutions of a second-order equation $Q_2(k)=0$,
where polynomial $Q_2(k)$ is the ratio of the polynomial appearing in Eq.~\eqref{k4} and of ${P_2(k)=k^2-k(k_2+k_1)+k_2k_1}$, i.e., explicitly

\[
k^4+ia_1k^3+a_2k^2+ia_3k+a_4=P_2(k)Q_2(k)\,.
\]

From here we find

\begin{widetext}
\be
Q_2(k)=(k+i\a_1)^2+k(k_2+k_1)+(2i\a_1+k_2+k_1)(k_2+k_1)+\a_2^2-\Delta-k_1k_2
\ee
\end{widetext}

and, hence,
\begin{subequations}
 \label{k34}
\bea
k_3=\frac{1}{2}\left[\mp i\sqrt{-\Delta_k^2-4\b^2\Delta_k/\Delta_p}+\sqrt{D_k}\right], \\
k_4=\frac{1}{2}\left[\mp i\sqrt{-\Delta_k^2-4\b^2\Delta_k/\Delta_p}-\sqrt{D_k}\right],
\eea
\end{subequations}
where $D_k=\Delta_k^2+4\b^2\frac{\Delta_p}{\Delta_k}+4k_2k_1$.
The sign before the first square root in Eqs.~\eqref{k34} should be chosen in accordance with the sign in Eqs.~\eqref{alp} for parameters $\a_{1,2}$.
To find $p_{3,4}$, we do not need to solve any equation.
We simply notice that the equation $\mbox{det}\tilde{F}(k)=0$ is invariant
under the transformation
$k \leftrightarrow p$, $\a_1 \leftrightarrow \a_2$,
$\Delta \leftrightarrow -\Delta$.
This means that, being transformed according to these rules,
Eqs.~\eqref{k34} give us the $p$ values:
\begin{subequations}
 \label{p34}
\bea
p_3 = \frac{1}{2}
\left[\mp i
\sqrt{-\Delta_p^2-4\b^2\Delta_p/\Delta_k}-\sqrt{D_p}\right], \\
p_4 = \frac{1}{2}
\left[\mp i
\sqrt{-\Delta_p^2-4\b^2\Delta_p/\Delta_k}+\sqrt{D_p}\right],
\eea
\end{subequations}
where $D_p = \Delta_p^2+4\b^2\frac{\Delta_k}{\Delta_p}+4p_2p_1$.

Let us now consider several important examples.
According to the analysis of Subsec.~\ref{subsec:zeros}, only two essentially different
possibilities exist, namely, a system described by the Cox
potential may have either one resonance or no resonance.

\subsubsection{One resonance}

Assume Eqs.~\eqref{sys} have two complex roots and let us define their first-channel components as
\begin{subequations}
\label{k12}
 \begin{eqnarray}
k_1 &=& k_r+ik_i, \\
k_2 &=& -k_r+ik_i,
 \end{eqnarray}
\end{subequations}
where $k_r$ and $k_i$ are real.
Let us assume the real part $k_r$ to be positive to fix ideas.
The imaginary part $k_i$, on the other hand, can be either positive or negative.
Let us write the corresponding energies, $k_{1,2}^2$, as $E_r \pm i E_i$,
where we also assume $E_i$ positive to fix ideas (which means that the upper sign corresponds to $k_1$ or $k_2$, depending on the sign of $k_i$).
We would like to choose as parameters the threshold difference $\Delta$,
as well as the real and imaginary parts of the resonance complex energy, $E_r, E_i$.
As seen below, these can correspond to physical parameters of a visible resonance in some (but not all) cases.
In terms of these parameters, $k_r$ and $k_i$ are expressed as
\begin{subequations}
\label{krootsri}
\bea \label{kroots} k_r&=& \frac{E_i}{\sqrt
2}\,\left[\sqrt{E_r^2+E_i^2}-E_r\right]^{-1/2}\,,\\
 k_i&=&\pm \frac{1}{\sqrt
2}\,\left[\sqrt{E_r^2+E_i^2}-E_r\right]^{1/2}\,.
\label{kroots2}
\eea
\end{subequations}
In the second channel,
the roots corresponding to $k_{1,2}$ can be found from the threshold condition~\eqref{threshold}.
They are given by
\begin{subequations}
 \begin{eqnarray}
p_1 &=& p_r+ip_i, \\
p_2 &=& -p_r+ip_i,
 \end{eqnarray}
\end{subequations}
with
\begin{subequations}
\bea\label{proots}
p_r&=& -\frac{1}{\sqrt
2}\,\left[\sqrt{E_i^2+(E_r-\Delta)^2}+E_r-\Delta\right]^{1/2}\!\!, \\
 p_i&=& \mp\frac{E_i}{\sqrt2}\,\left[\sqrt{E_i^2+(E_r-\Delta)^2}+E_r-\Delta\right]^{-1/2}\!\!\!\!\!\!.
\label{proots2}
\eea
\end{subequations}
The upper (resp., lower) sign in Eq.~\eqref{kroots2} corresponds to the upper (resp., lower) sign in Eq.~\eqref{proots2}, which means that, for a given zero, the signs of $k_i$ and $p_i$ are opposite.
This can be understood from the first column of Fig.~\ref{fig1}:
a resonance appears when the two hyperbolas are tangent to each other, which can only happen in the second and fourth quadrant, where $\lambda$ and $\rho$ have opposite signs.
Moreover, Eqs.~\eqref{kroots} and~\eqref{proots} show that, for a given zero, the signs of $k_r$ and $p_r$ are also opposite.
This implies that, for the Cox potential, the complex resonance zeros (or scattering-matrix poles) are always in opposite quadrants in the complex $k$ and $p$ planes, as illustrated for instance by the complex zeros in Fig.~\ref{fig1}(c).

This has important consequences for physical applications:
for a resonance to be visible, one of the corresponding zero has to lie close to the physical positive-energy region, i.e., close to the real positive $k$ axis and close to the region made of the real positive $p$ axis and of the positive imaginary $p$ interval: $[0,i\sqrt{\Delta}]$.
Consequently, the only possibility for a visible resonance with the Cox potential is that of a Feshbach resonance, only visible in the channel with lowest threshold, with an energy lying below the first threshold.
At higher resonance energies, the corresponding zero is either close to the $k$-plane physical region (and far from the $p$-plane one) or close to the $p$-plane physical region (and far from the $k$-plane one); it cannot be close to both physical regions at the same time, hence it cannot have a visible impact on the coupled scattering matrix.
Here, we limit ourselves to the case of a visible resonance, which is the most interesting from the physical point of view.
It corresponds to the lower signs in Eqs.~\eqref{kroots2} and~\eqref{proots2},
with a resonance energy $E_r$ such that $0<E_r<\Delta$, and a resonance width $\Gamma=2 E_i$ such that $E_i<E_r$.

To get a potential without bound state,
we choose the upper signs in Eqs.~\eqref{alp}, which leads to
\begin{subequations}
\label{a12}
\bea
 \a_{1}&=&-k_i+ k_r\left[ -1-\b^2/(k_rp_r)\right]^{1/2}\,, \\
 \a_{2}&=&-p_i- p_r\left[ -1-\b^2/(k_rp_r) \right]^{1/2}\,.
\eea
\end{subequations}

\begin{widetext}
\begin{figure}[ht]
\begin{center}
\begin{minipage}{17.8cm}
\epsfig{file=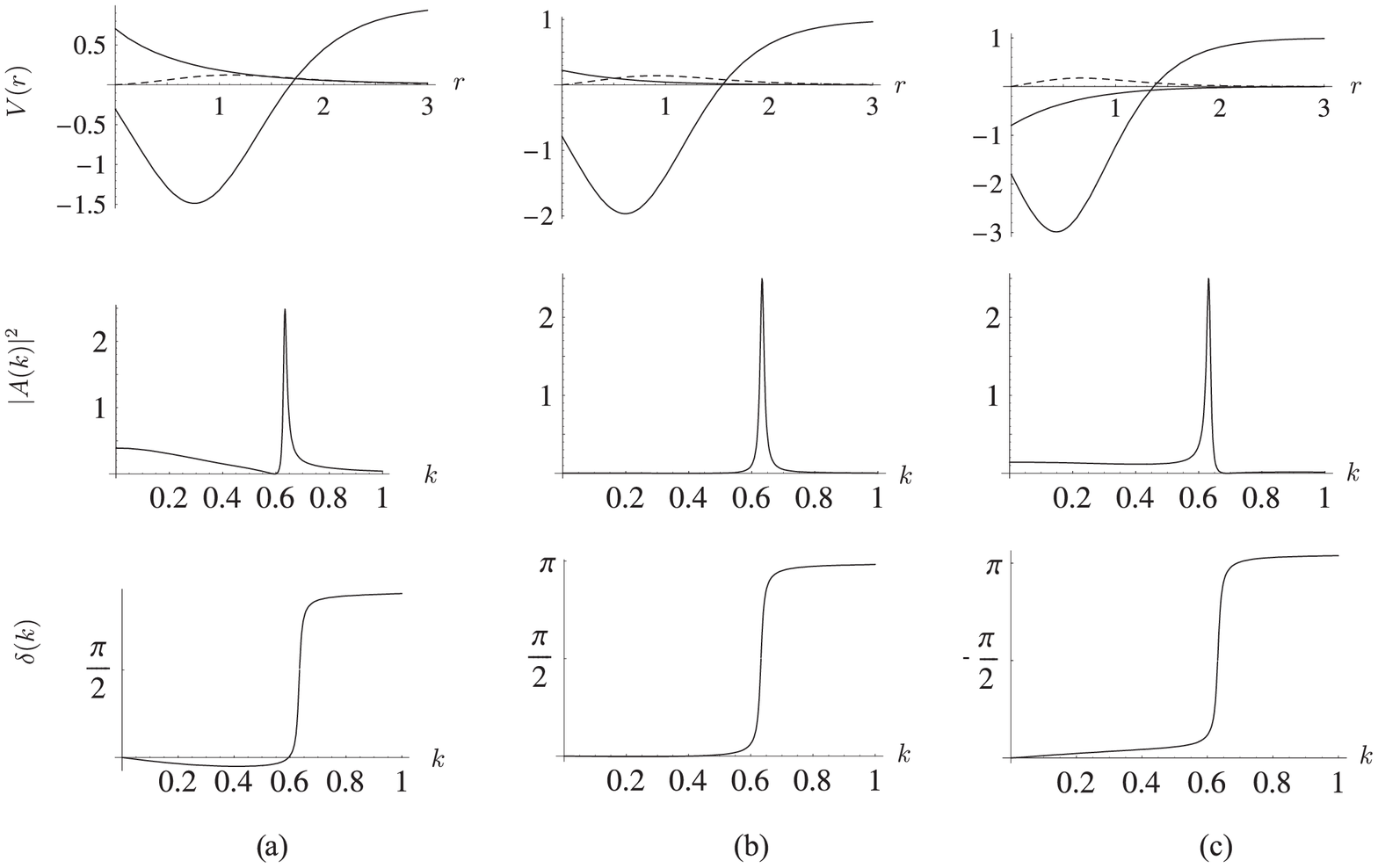, width=17cm}
\caption{\small The Cox potential with one visible resonance of energy
$E_r=0.4$ and width $\Gamma=0.02$, for $\Delta=1$ and $\beta=0.1$
(first row, solid lines for $\tilde{v}_{11}$ and $\tilde{v}_{22}+\Delta$,
dashed line for $\tilde{v}_{12}$), with the corresponding partial cross section (second row)
and phase shifts (third row) for (a) $\kappa_1 = 0.5$; (b) $\kappa_1=0.7$; (c) $\kappa_1 = 1$. \label{figExR}}
\end{minipage}
\end{center}
\end{figure}
\end{widetext}

From here, we see that,
for non-zero values of the parameters $k_r$ and $p_r$ (which have opposite signs),
the coupling parameter $\b$ cannot be infinitesimal:
because $\a_{1}$ and $\a_{2}$ have to be real,
$\b$ is restricted to satisfy the inequality $\b \ge \sqrt{-k_r p_r}$.
Now, using Eqs.~\eqref{k34} and~\eqref{p34} it is easy to find the two
remaining roots
\begin{subequations}
\be\label{k11} k_{3,4}=-i\sqrt{-k_r^2-\b^2k_r/p_r}\pm
i\sqrt{k_i^2-\b^2p_r/k_r}\,,
\ee
\be\label{k21}
p_{3,4}=-i\sqrt{-p_r^2-\b^2p_r/k_r}\mp i\sqrt{p_i^2-\b^2k_r/p_r}\,.
\ee
\end{subequations}

To get a potential with one bound state at energy $-\lambda_b^2$,
we choose the lower signs in Eqs.~\eqref{alp}.
We then get for $k_3(\beta)$ an expression similar to Eq.~\eqref{k11},
from which the value of $\beta$ can be found by solving the bi-squared equation
\be
k_3(\b)=i \lambda_b\,.
\ee

Let us now choose explicit parameters.
First, without loosing generality we may put $\Delta=1$ (see Subsec.~\ref{subsec:zeros}).
To get a visible resonance, we put $E_r=0.4$, $E_i=0.01$ (which corresponds to a resonance width $\Gamma=0.02$), and $\b=0.1$.
Using Eqs.~\eqref{krootsri} and~\eqref{a12}, one finds $\a_1=0.76938$ and $\a_2=-0.766853$.
The factorization energy, $\cal E$, is not constrained in this case:
it just has to be negative.
The Cox potential with one resonance and no bound state for different values of $\kappa_1$
is shown in the first row of Fig.~\ref{figExR}.
The diagonal elements of the potentials, $V_{11}$ and $V_{22}+\Delta$,
are plotted with solid lines,
while $V_{12}$ is plotted with dashed lines.
Parameter $\kappa_1$ is responsible for changing the range of the potential,
as shown by Eqs.~\eqref{vtcox},
and hence the scattering length (see discussion below).
The second row of this figure shows the corresponding partial cross sections,
where the resonance behavior is clearly seen, as well as the evolution of the low-energy cross section, which is related to the scattering length.
The last row of Fig.~\ref{figExR} shows the corresponding phase shifts for the open channel,
where a typical Breit-Wigner behavior (see e.g.\ Ref.~\cite{taylor:72}) is seen for the resonance, as well as the evolution of the zero-energy phase-shift slope, which is also related to the scattering length.
\subsubsection{Two bound states}

\begin{figure}
\begin{center}
\begin{minipage}{8.6cm}
\epsfig{file=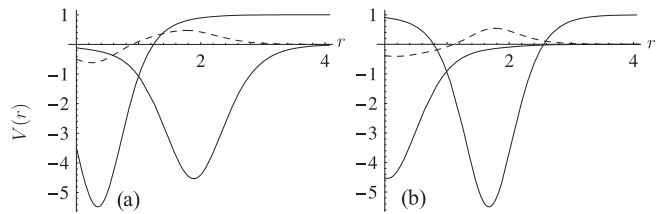, width=8.6cm}
\caption{\small The Cox potential (solid lines for $\tilde{v}_{11}$ and $\tilde{v}_{22}+\Delta$, dashed line for $\tilde{v}_{12}$) with two bound states at energies $E_1=-0.01$ and $E_2=-2.25$, for $\Delta=1$, $\beta=0.1$ and $\kappa_1 = 1.51$.
The left (resp., right) graphic corresponds to the upper (resp., lower) signs
 in Eqs.~\eqref{alp}.  \label{figExB2}}
\end{minipage}
\end{center}
\end{figure}

Let us now construct a Cox potential with two bound states,
and hence no resonance (see Fig.~\ref{fig2}).
We choose $k_1=0.1i$ and $k_2=1.5i$ for these bound states and,
as in the previous example, we put $\Delta=1$ and $\b=0.1$.
We thus have $p_1=\sqrt{1.01} i$ and $p_2=\sqrt{3.25} i$,
which defines $\Delta_p$ in Eqs.~\eqref{alp}.
Choosing the upper signs in these equations,
we find $\a_1=-0.112649$ and $\a_2=-1.79557$,
while for the lower signs, we get $\a_1=-1.48735$ and $\a_2=-1.0122$.
The corresponding Cox potentials are shown in Fig.~\ref{figExB2}.

\subsubsection{One bound state}

\begin{figure}
\begin{center}
\begin{minipage}{8.6cm}
\epsfig{file=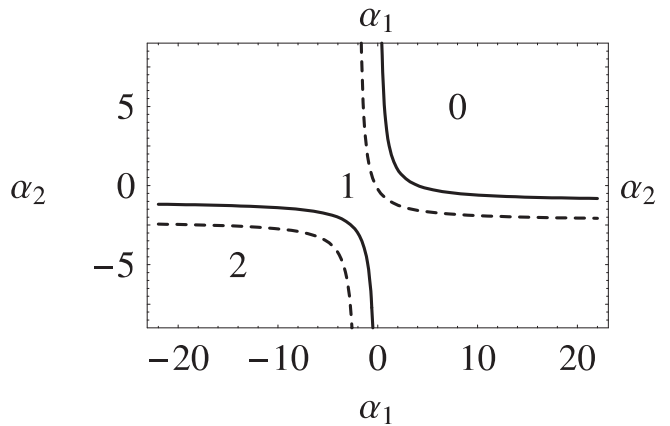, width=8.6cm}
\caption{\small Iso-energy curves defined by Eq.~\eqref{c1} (dashed lines) in the plane of parameters $(\a_1,\a_2)$ for $E_b=-4$, $\Delta=1$, $\b=2$, together with the curves defined by Eq.~\eqref{bsz} (solid lines).
\label{figex4}}
\end{minipage}
\end{center}
\end{figure}

If only one bound state $E_b=-\lambda_b^2$ is known,
we may consider the position of the second zero as a free parameter.
Besides, one can solve system~\eqref{syslr} at $\lambda=\lambda_b$
with respect to $\a_2$, which leads to
\be\label{c1}
\a_2=\frac{\b^2}{\lambda_b+\a_1}-\sqrt{\Delta+\lambda_b^2}.
\ee
Figure~\ref{figex4} gives a graphical representation of Eq.~\eqref{c1} in the plane of parameters $(\a_1,\a_2)$:
the dashed curves correspond to all the Cox potentials that have a bound state at the same energy $E_b=-4$.
Figure~\ref{figex4} also shows that these iso-energy curves~\eqref{c1}
can be obtained by a shifting of the curves separating regions of plane $\mathbb{A}$ with different number of bound states, defined by Eq.~\eqref{bsz} (solid lines).

Let us finally summarize these possible inverse problems in Table \ref{tab2}.
The free parameters allow either for isospectral deformations of the potential or for fits of additional experimental data as, e.g., scattering lengths.
This possibility will be used in the next section on atom-atom systems.
\begin{center}
\begin{table}\caption{\label{tab2}
Possible mappings between some experimental data
and the Cox potential parameters.}
\begin{tabular}{|l||l||l||l|}
\hline Experimental & Fixed  & Free  & Restrictions  \\
data & parameters  & parameters &
\\
 \hline
 $\Delta\,, E_r\,, E_i$ & $\a_1\,, \a_2\, $   &
 $\kappa_1, \b$ & $\b\geq \sqrt{-k_rp_r}$
\\ \hline
$\Delta\,, E_b=-\lambda_b^2\,, E_r\,, E_i $ &$\a_1\,, \a_2\,, \b$
 & $\kappa_1$ & $\kappa_1>\lambda_b$
\\ \hline
 $\Delta\,, E_{1,2}=-\lambda_{1,2}^2$ & $\a_1\,, \a_2\, $   &
 $\kappa_1, \b$ & $\kappa_1>\lambda_2>\lambda_1$
\\ \hline
 $\Delta\,, E_{b}=-\lambda_b^2 $ &  $\a_2\, $ & $\kappa_1\,, \b\,, \alpha_1$ &
 $\kappa_1>\lambda_b$
\\ \hline
\end{tabular}
\end{table}
\end{center}

\subsection{Low-energy scattering matrix}

In this section,
we analyze the $S$-matrix given by Eq.~\eqref{S}
for energies close to the lowest threshold,
the energy of which we have chosen equal to zero.
From Eqs.~\eqref{F} and~\eqref{det1}, one finds the Cox-potential $S$-matrix
\be\label{sm}
\mathbf{S}(k,p)= \frac{1}{f(k,p)} \left(\begin{array}{cc}
f(-k,p) & \frac{-2i\b\sqrt{kp}}{k^2+\kappa_1^2}
\\[.25em]
\frac{-2i\b\sqrt{kp}}{p^2+\kappa_2^2} & f(k,-p)
\end{array}
\right).
 \ee
When the second channel is closed, i.e., for energies $0<E<\Delta$,
the physical scattering matrix is just a function $S(k,p)$,
which coincides with the first diagonal element of $S$-matrix~\eqref{sm}.
It reads
\be\label{sf}
S(k,p)=\frac{k+i\kappa_1}{k-i\kappa_1} \frac{[i(k-i\a_1)
(\sqrt{\Delta-k^2}+\a_2)-\b^2]}{[i(k+i\a_1)
(\sqrt{\Delta-k^2}+\a_2)+\b^2]}\,. \ee
From here one finds the
scattering amplitude ${A(k)=[S(k)-1]/2ik}$, which reads
\be A(k)=
\frac{(\a_2+\sqrt{\Delta-k^2}\,\,)
\left(\a_1-\kappa_1\right)-\b^2}{i\left(k-i\kappa_1\right)
\left[i(k+i\a_1)\left(\sqrt{\Delta-k^2}+\a_2\right)+\b^2\right]}\,,
\ee
and the scattering length $a=-A(0)$, which reads
\be\label{sl}
a=
\frac{1}{\kappa_1}+\frac{\sqrt{\Delta}+\a_2}{\b^2-\a_1\left(\sqrt{\Delta}+\a_2\right)}\,.
\ee
From the argument of $S(k)=e^{2i\d (k)}$, one deduces the
phase shift $\d (k)$, which reads
\be\label{ps} \d (k)=\arctan
\frac{k}{\kappa_1} +\arctan
\frac{k\left(\sqrt{\Delta-k^2}+\a_2\right)}{\beta^2-\alpha_1\left(\sqrt{\Delta-k^2}+\a_2\right)}\,.
\ee
One can check on Eqs.~\eqref{sl} and~\eqref{ps} that the
scattering length is the slope of the phase shift at zero energy,
as it should be. Note that Eq.~\eqref{ps} is equivalent to \be
k\cot\d(k)=\frac{a_{\b}(k)\kappa_1+k^2}{\kappa_1-a_{\b}}\,, \ee
where $a_{\b}(k)=\a_1-\b^2/\left(\sqrt{\Delta-k^2}+\a_2\right)$.
In the uncoupled case ($\b=0$), this expression reduces to the
phase shifts of the simplest
 Bargmann potential (see e.g.\ Ref.~\cite{newton:82}),
 which depends on the parameters $\kappa_1$ and $a_B \equiv a_{\beta=0}=\alpha_1$.
Therefore, the Cox potential may be considered as a
coupled-channel deformation of
the Bargmann potential, resulting in an energy dependence
of one of its parameters, $a_B$.

The scattering length is an important physical quantity.
In many-body theories for instance,
it is often used to describe interactions in the $s$-wave regime.
Let us thus study into detail the scattering length of the Cox potential,
as given by Eq.~\eqref{sl}.
When considered as a function of $\a_{1,2}$,
it has a singularity located at the boundary of
the single-bound-state region provided by Eq.~\eqref{bsz}.
Such infinite values of the scattering length happen when a zero of the Jost determinant, which corresponds to an $S$-matrix pole, crosses the first threshold: a bound state is then transformed into a virtual state,
in agreement with the general theory~\cite{newton:82}.
We can analyze the sign of the scattering length
for different numbers of bound states $n_b$ given by Eq.~\eqref{nb},
by considering the indices $I_{1,2}$ given by Eqs.~\eqref{ind12}.
First, we remark that $I_1=-1$ for $n_b=2,0$,
and $I_1=1$ for $n_b=1$.
Then, since $I_2=-1$ for $n_b=2$, the scattering length
is positive when the Cox potential has two bound states.
Next, for $n_b=0$ we have $I_2=1$;
the two contributions of the scattering length then have different
signs.
As a result, at fixed $\a_{1}$,  $\a_{2}$, $\b$, and $\Delta$,
one can get both positive and negative scattering lengths by varying
only $\kappa_1$.
Similarly, when $I_2=1$ (one-bound state),
the scattering length may only be positive.
In contrast, for $I_2=-1$ it may be both positive and negative but
for fixed values of $\a_1$, $\a_2$, $\b$, and $\Delta$, it becomes
negative for large enough $\kappa_1$.

\section{Two-channel model of alkali-metal atom-atom collisions in the presence of a magnetic field \label{4}}

\subsection{Magnetic Feshbach resonance}

Ultra-cold collisions of alkali-metal atoms play a key role in applications of laser
cooling such as Bose-Einstein condensation and BEC-BCS crossover. The analysis of such
experiments is commonly based on the coupled-channel method~\cite{stoof:88}, i.e., on
solving numerically a set of coupled differential equations.

In this paper, we reduce the
low-energy scattering problem of two
alkali-metal atoms to an effective two-channel problem with a
single Feshbach resonance, as in Ref.~\cite{nygaard:06}.
The model consists of a single closed channel $Q$ containing a bound state,
which interacts with the scattering continuum in
the open channel $P$,
so that the whole scattering problem is reduced to the two-channel scattering described by the $2\times2$ Hamiltonian
\be\label{HPQ}
H=-\frac{d^2}{dr^2}+
\left(
\begin{array}{cc}
 V_{P}(r) & V_{int}(r) \\[.3em]  V_{int}(r)&
 V_{Q}(r)
\end{array}\right),
\ee
where $V_{P}$ is the uncoupled open-channel potential,
$V_{Q}$ is the uncoupled closed-channel potential,
and potential $V_{int}$ describes the coupling between the open and
closed channels $P$ and $Q$.
These channels describe atoms placed in a magnetic field and
occupying different energy sub-levels
which can be shifted with respect to each other
with the change of the magnetic field (Zeeman effect).
For each value of the magnetic field,
the zero of energy is chosen as the energy
of the dissociated atoms in channel $P$.

Even in the simplest case of a homogeneous magnetic field,
the potential-energy matrix of Hamiltonian~\eqref{HPQ} depends on
the magnetic field.
We will assume that the external field changes slowly enough
so that we can take advantage of the adiabatic approximation,
assuming that the stationary Schr\"odinger equation
may be applied for describing the scattering process and the
magnetic field enters the Hamiltonian as a parameter only.
Moreover, the known observation that,
when the scattering length is much larger than the range of the interaction,
the general behavior of the system is nearly independent
of the exact form of the potential~\cite{sakurai:94},
suggests us to use the Cox potential with large scattering length for
describing the interatomic scattering.
We thus replace the potential matrix in Eq.~\eqref{HPQ} by the Cox potential.
In this case, the parameters of the Cox potential
should carry a dependence on the magnetic field.
Below, we show that,
to get a good agreement with available experimental data,
it is sufficient to impose a linear field dependence on the threshold difference $\Delta$ only,
keeping all other parameters field independent.
Thus, inverting known scattering experimental data,
one can find all the parameters defining the Cox potential, obtaining
in this way a simple analytical model of the atom-atom scattering process in
the presence of a magnetic field.

The position of the highest bound (or virtual) state
is crucial
 in describing the resonance phenomena
of interatomic collisions.
In an $s$-wave single-channel system, the scattering
process becomes resonant at low energy when a bound state or virtual state is
located near the threshold,
a phenomenon known as ``potential resonance''.
In a multichannel system, the incoming channel
(which is always open) may be coupled during the collision
process
to other open or closed channels, corresponding to different
spin configurations.
When a bound state in a closed channel
lies near the collision energy continuum, a Feshbach
resonance~\cite{feshbach:58,feshbach:62} may occur, giving rise to scattering properties
that are tunable by an external magnetic field.
In Ref.~\cite{marcelis:04}, some interesting examples of the interplay
between a potential resonance and a Feshbach resonance are considered.
Below, we adjust the analytically-solvable model based on the
Cox potential for describing the same phenomena.

Typically, the coupling between the closed and open channels is
rather small; we thus consider first an uncoupled limit of the Cox potential,
i.e., $V_{int}(r)\to0$, which  corresponds to $\b\to0$.
In this case, the Jost determinant~\eqref{det1} has the following zeros:
\be
k_1=-i\a_1
\ee
and
\be
p_2=-i\a_2.
\ee
The energies of these unperturbed
(i.e., with zero coupling)
 states
(called bare molecular states in Ref.~\cite{marcelis:04}) are
\be\label{bme}
E_1=-\a_1^2
\ee
and
\be
E_2=-\a_2^2+\Delta.
\ee
It should be noted that
in this case
$E_1$ belongs to channel $P$ while
$E_2$ belongs to channel $Q$.
Hence, $\a_1$ is associated with the potential resonance,
while $\a_2$ is associated with the Feshbach resonance.
Due to the Zeeman effect, the difference between the thresholds
is a linear function of the magnetic
field,
\be
 \Delta(B)=\Delta_0+\mu_{mag}(B-B_0),
 \label{DeltaB}
 \ee
where $B_0$ can be arbitrarily chosen in the domain of interest and $\Delta_0$ is the value of the threshold corresponding to $B_0$.
If $\a_{1,2}<0$ and the coupling is absent,
then the two bound states
cross at $\Delta=\a_2^2-\a_1^2$.
Note that $E_2$ crosses the threshold at $\Delta=\a_2^2$.
When there is a coupling between channels,
the levels $E_1$ and $E_2$ avoid crossing (see below).

Let us consider the behavior of the scattering length in the presence of the Feshbach resonance.
It is described by the following formula~\cite{moerdijk:95}:
\be\label{slt}
a=a_{bg}\left(1-\frac{\Gamma_B}{B-B_0}\right).
\ee
Here, $B_0$ is the position of the magnetic Feshbach resonance and $\Gamma_B$ is its width (in terms of magnetic field).

In particular, Eq.~\eqref{sl} shows that such an infinite
value of the scattering length occurs for the Cox potential at a
threshold $\Delta_0$ defined by:
\begin{equation}
\sqrt{\Delta_0}=\frac{\b^2-\a_1\a_2}{\a_1}\,.
\end{equation}
Let us now assume for the Cox potential a threshold difference given by
Eq.~\eqref{DeltaB}  with such a value of $\Delta_0$. Expanding Eq.~\eqref{sl} near this
resonance, it is easy to get Eq.~\eqref{slt} and a
simple expression for the width of the resonance for the Cox potential.


\begin{eqnarray}\label{CoxSLFr}
a &=& \frac{\a_1-\kappa_1}{\a_1\kappa_1}\\
& \times & \left(1+\frac{2\left[1+{\rm o}\left(\Delta-\Delta_0\right)\right]\kappa_1\sqrt{\mathstrut \Delta_0}\left(\sqrt{\mathstrut \Delta_0}+\a_2\right)}
{\left(\a_1-\kappa_1\right)\left(\Delta_0-\Delta \right)}\right)\,.\nonumber
\end{eqnarray}

\be
\Gamma_B=\frac{2\kappa_1\sqrt{\mathstrut \Delta_0}\left(\sqrt{\mathstrut \Delta_0}+\a_2\right)}
{\mu_{mag}\left(\a_1-\kappa_1\right)}\,.
\ee

As shown in Ref.~\cite{marcelis:04}, the background scattering length $a_{bg}$
is due to the open-channel potential.
When there is a bound state or virtual state close to threshold,
it can be further decomposed as a sum of two contributions:
a standard potential part, which depends on the potential range,
and a potential-resonance part, which depends on the bound/virtual-state energy.
This decompostion clearly appears in our model:
the background scattering length corresponds to a large magnetic field $B$ in Eq.~\eqref{slt} or to a large $\Delta$ in Eq.~\eqref{sl},
namely, $a_{bg}=\lim\limits_{\Delta\rightarrow\infty}a$, which yields:
\be\label{abg}
a_{bg}=\frac{1}{\kappa_1}-\frac{1}{\a_1}.
\ee
The same result comes from Eq.~\eqref{CoxSLFr}. It should be noted that a large threshold difference
$\Delta\rightarrow\infty$ effectively corresponds to a small coupling $\b\rightarrow 0$. From here we find
another relation $a_{bg}=\lim\limits_{\b\rightarrow 0}a$ which leads to the same background scattering length.

In this formula,
the first term is proportional to $1/\kappa_1$, the parameter
which defines the range of the open-channel potential [see Eqs.~\eqref{vtcox}];
it may thus be considered as the standard potential part of the background scattering length.
The second term is associated with the $P$-channel bound (or virtual)
state in the uncoupled limit.
Hence, it may be interpreted as the potential-resonance part of the background scattering length.
Let us further consider two different possibilities giving rise to a
large (either positive or negative) background scattering length.

\subsection{Interplay between a bound state and the Feshbach resonance}

The first possibility occurs when the highest bound state is located near
the threshold, i.e., when $\a_1\lesssim 0$.
\begin{figure}
\begin{center}
\epsfig{file=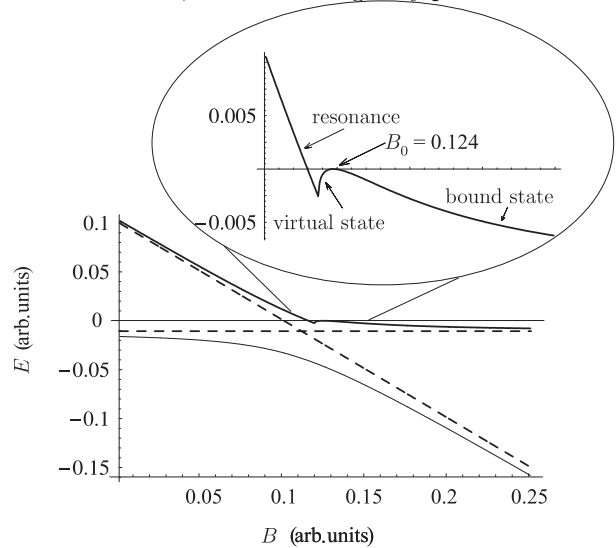, width=8cm}
\caption{\small Energies of bare
(dashed lines) and dressed (solid lines) states
as functions of the magnetic field $B$ for the Cox potential defined by parameters~\eqref{parCs}.
The transition between a Feshbach resonance, a virtual state, and a bound state is shown in the inset.
\label{figE}}
\end{center}
\end{figure}
In Fig.~\ref{figE}, we show energies as functions of the magnetic field
when channel $P$ has a bound state just below the threshold, for
\begin{subequations}
\label{parCs}
\begin{eqnarray}
\b & = & 0.05, \\
\a_1 & = & -\lambda_b=-0.103, \\
\a_2 & = & -0.5, \\
\kappa_1 & = & 1. \label{kappa1}
\end{eqnarray}
\end{subequations}

\begin{figure}
\begin{center}
\epsfig{file=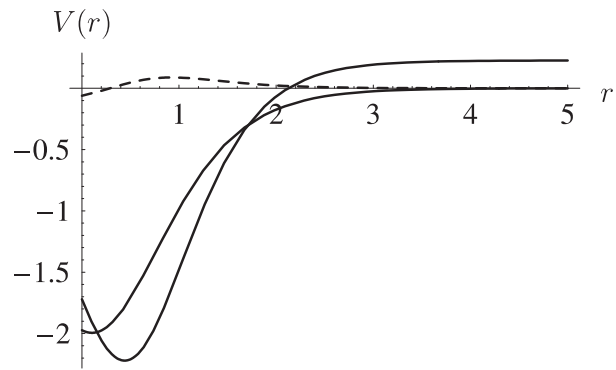, width=8cm}
\caption{\small The Cox potential defined by parameters~\eqref{parCs} for $B=0.1$;
$V_{P}$ and $V_{Q}+\Delta$ are represented by solid lines,
$V_{int}$ by a dashed line.
\label{figptCs}}
\end{center}
\end{figure}

We are using arbitrary units and choose $\Delta(B)=0.35-B$ in Eq.~\eqref{DeltaB}.
The bare bound states ($\beta=0$) of the $P$ and $Q$ channels are indicated by the dashed horizontal and
slanted lines respectively.
These bare states are Hamiltonian eigenstates in the uncoupled $P$ and $Q$ subspaces.
The bare $Q$-channel bound state crosses the $P$-channel threshold (solid
horizontal line) at $B=0.1$.
The dressed states are represented by solid lines and display an avoided-crossing behavior~\cite{marcelis:04}.
For low fields, the model has one bound state and one Feshbach resonance,
the energies of which are close to the bare-state energies.
The energy of the Feshbach resonance becomes negative above $B=0.112$,
when the imaginary part of the zero in the complex $k_1$ plane becomes larger than its real part.
At $B=0.12$, these complex zeros collapse and transform into two virtual states (purely imaginary zeros), which corresponds to the discontinuous slope in Fig.~\ref{figE}.
With increasing magnetic field, one of these virtual states (represented in Fig.~\ref{figE}) gets closer to threshold, while the other one (not represented in Fig.~\ref{figE} as it does not affect the low-energy scattering properties) goes away.
At $B_0=0.124$, the virtual state crosses the threshold and becomes a bound state;
the scattering length thus goes through infinite values at that field: this is the magnetic-Feshbach-resonance phenomenon itself.
\begin{widetext}
\begin{figure}[ht]
\begin{center}
\begin{minipage}{17cm}
\epsfig{file=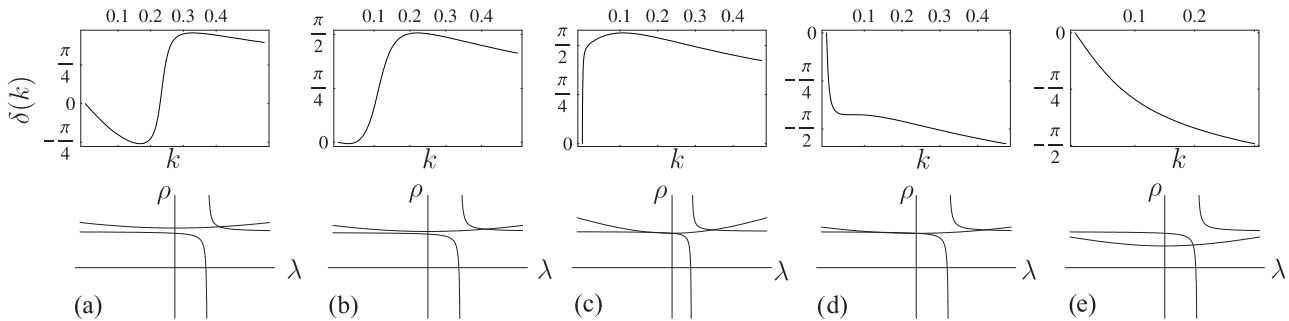, width=17cm} 
\caption{\small Phase shifts and graphical representation of Eqs.~\eqref{syslr}
for the Cox potential defined by parameters~\eqref{parCs}.
The columns correspond to different values of the magnetic field:
(a) $B = 0.05$; (b) 0.1; (c) 0.1235; (d) 0.125; (e) 0.24.
\label{figSH}}
\end{minipage}
\end{center}
\end{figure}
\end{widetext}
Above $B_0$, the model has two bound states, the energies of which tend to the bare-state energies when the field increases.

Following Ref.~\cite{marcelis:04},
we stress that, although the
behavior of the dressed states shows some resemblance with the two-level Landau-Zener description, this model does not include the threshold effects shown in Fig.~\ref{figE}
and, hence, cannot be used to properly
describe the interplay between a potential resonance and a Feshbach resonance.
With respect to Ref.~\cite{marcelis:04},
our model displays a slightly more sophisticated behavior for the state energies (compare our Fig.~\ref{figE} with their Fig.~4).
A more significant novelty of our description is the direct knowledge of the coupled-channel potential corresponding to these energies.
This potential is shown in Fig.~\ref{figptCs} for $B=0.1$.
The potential form factor changes slowly with the change of the magnetic field,
which is mainly responsible for the variation of $\Delta$.

The value of $\kappa_1$ chosen in Eq.~\eqref{kappa1} is arbitrary.
However, the necessary and sufficient condition to get a Cox potential without singularity imposes then that the bound-state energies of the model should be larger than $-1$.
Figure~\ref{figE} shows that this condition will be satisfied for a limited range of magnetic field only.
For higher fields, a larger $\kappa_1$ should be chosen.
The phase shifts of the same Cox potential,
as well as a graphical representation of Eqs.~\eqref{syslr},
are shown in Fig.~\ref{figSH} for different values of $B$.
The first and the last columns correspond to a large positive background scattering length
($a_{bg}\sim1/\lambda_b\approx 10$),
due to a bound state close to the threshold.
Physically, this occurs for the $^{133}\rm{Cs}$ atom-atom interaction~\cite{leo:00}, for instance.
Figure~\ref{figSH}(b) illustrates the case where the scattering length is close to zero.
The calculation or measurement of the zero of the
scattering length plays an important role in determining the resonance width~\cite{ohara:02}.
The phase-shift behavior for the virtual state and bound state close to threshold is shown in Figs.~\ref{figSH}(c) and \ref{figSH}(d), respectively.
In this case, the scattering length is very large
and its sign changes while the energy of the zero of the Jost-matrix determinant crosses the threshold.
Recalling that the intersection points in the graphical representation of
Eqs.~\eqref{syslr}, shown in the second row of Fig.~\ref{figSH}, give
the positions of bound and virtual states,
one may establish a correspondence between the second row of Fig.~\ref{figSH}
and the motion of the corresponding zeros in the complex plane described above.

\subsection{Interplay between a virtual state and the Feshbach resonance}

Another interesting possibility occurs when there is a virtual state close
to the threshold, i.e., when $\a_1\gtrsim 0$.
This is the case of the $^{85}{\rm Rb}$ atom-atom interaction, for example.
We will use rubidium scattering data~\cite{arimondo:77,marcelis:04} in this example,
and work with units $\hbar=2\mu=1$, where $\mu$ is the reduced mass of the two atoms.
The length unit is chosen as the Bohr radius $a_0$;
energies are thus expressed in units of $a_0^{-2}$.
According to Ref.~\cite{marcelis:04}, the bare virtual
state is located at $\lambda_v=-1.78\cdot10^{-3}a_0^{-1}$, but this value is associated with the model they used in their calculations.
We just consider $\lambda_v\sim -10^{-3}a_0^{-1}$ and set Eq.~\eqref{abg} as a constraint between $\alpha_1=-\lambda_v$ and $\kappa_1$.
In order to fit the scattering-length behavior~\eqref{slt} with $a_{bg}=-443\,a_0$, $B_0=15.5041$ mT and $\Gamma_B=1.071$ mT, we use Eq.~\eqref{sl}.

The value of $\b$ defines, in particular,
the position of the Feshbach resonance,
i.e., the magnetic field $B_0$ for which the bound state crosses the threshold.
According to Eq.~\eqref{bsz}, one has
\begin{equation}
\b=\sqrt{\a_1\left(\a_2+\sqrt{\Delta_0}\right)},
\end{equation}
where $\Delta_0$ is the value of the threshold corresponding to $B_0$.
The value of $\a_2$, defining the width of the Feshbach resonance $\Gamma_B$,
should be found from the condition $a(B_0+\Gamma_B)=0$.
Then, according to Eq.~\eqref{sl}, we find
\be
\a_2=\frac{\a_1\left[\sqrt{\Delta(B_0+\Gamma_B)}-\sqrt{\Delta_0}\right]}{\kappa_1}
-\sqrt{\Delta(B_0+\Gamma_B)},
\ee

\begin{widetext}
\begin{figure}
\begin{center}
\begin{minipage}{17cm}
\epsfig{file=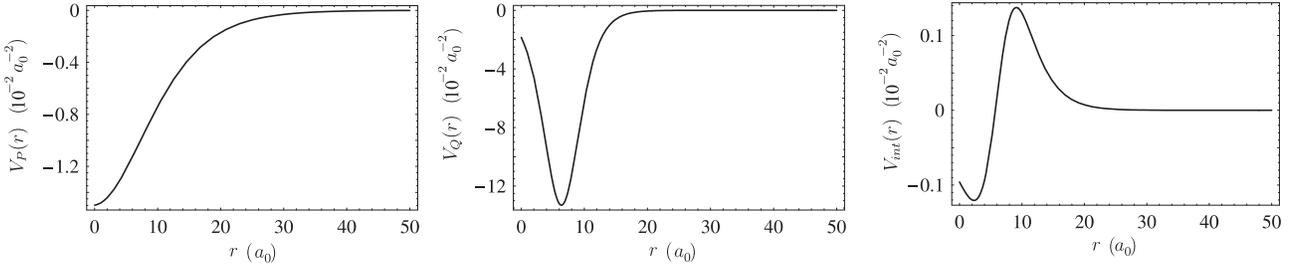, width=17cm}
\caption{\small The Cox potential describing the Feshbach resonance in $^{85}$Rb,
defined by parameters~\eqref{parRb}, plotted at $B=14.5$ mT ($\Delta=0.0590363\,a_0^{-2}$).
\label{RB-Cox}}
\end{minipage}
\end{center}
\end{figure}
\end{widetext}
where
$\Delta_0=2471.386$~MHz and $\mu_{mag}=-36.4$~MHz/mT~\cite{marcelis:04}.
To get that value of $\Delta_0$, we have used the known value of the threshold at zero magnetic field \cite{arimondo:77} and assumed that Eq.~\eqref{DeltaB} is valid down to that field.

From Eq.~\eqref{abg}, we may fix $\kappa_1=\a_1/(1+\a_{bg}\kappa_1)$ at $a_{bg}=-443\,a_0$
and find
the values of all parameters defining the potential at the given
position of the
Feshbach resonance and with the given value of the background scattering length:
\begin{subequations}
\label{parRb}
\bea
\!\!\!\!\!\! \b & = &0.0202366\,a_0^{-1}, \\
\!\!\!\!\!\! \a_1 & = &-\lambda_v=2.2\cdot10^{-3}\,a_0^{-1}, \\
\!\!\!\!\!\! \a_2 & = & -0.239343\,a_0^{-1}, \\
\!\!\!\!\!\! \kappa_1 & = & 0.0866\,a_0^{-1}, \label{kappa1Rb}\\
\!\!\!\!\!\! \kappa_2 & = & \sqrt{\kappa_1^2+\Delta}=\sqrt{0.0789668-0.856899B}\,a_0^{-1}.
\eea
\end{subequations}

\begin{figure}
\begin{center}
\begin{minipage}{8.5cm}
\epsfig{file=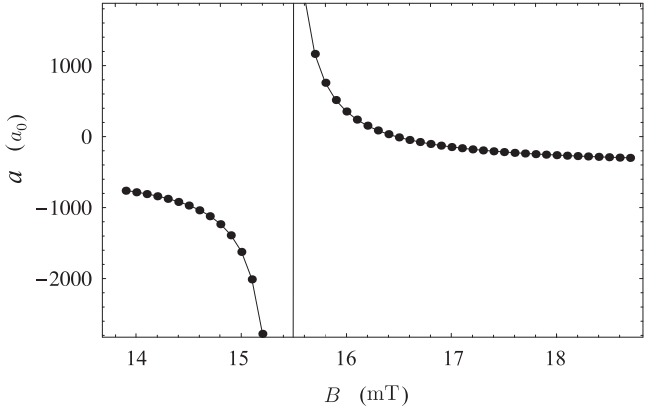, width=8.5cm}
\caption{\small Solid line: Feshbach-resonance scattering length~\eqref{slt}.
Dots: Cox-potential scattering length~\eqref{sl} for the parameters~\eqref{parRb}.
\label{figSl}}
\end{minipage}
\end{center}
\end{figure}

In Fig.~\ref{figSl}, we show that, with these parameters,
the Cox-potential scattering length~\eqref{sl} reproduces
the Feshbach-resonance scattering length~\eqref{slt}
with good precision.
The value $\a_1=2.2\cdot10^{-3}\,a_0^{-1}$ was chosen
to get a smooth potential $V_P$ without repulsive core.
This potential is shown in Fig.~\ref{RB-Cox} and, once again,
has a form factor rather independent of the field, except for the threshold.
In Fig.~\ref{figEV}, we show the corresponding energies as functions of the magnetic field.
The bare bound state of channel $Q$ is represented by the slanted dashed line.
The bare virtual state of channel $P$, which is located at $\lambda_v=-2.2\cdot10^{-3} \,a_0^{-1}$, is not shown in Fig.~\ref{figEV}.
The dressed states are indicated by solid lines.
When $B<B_0=15.5041$~mT, there exist both a virtual state and a Feshbach resonance,
the energies of which tend to the bare-state energies for small $B$.
The virtual state becomes a bound state at $B=B_0$ (see inset).
With increasing $B$, the real part of the resonance energy decreases
and at $B=16.657$~mT it crosses the threshold.
Finally, at $B=16.9$~mT, the two resonance poles collapse and produce two virtual states,
one of which stabilizes at $\lambda_v=-2.2\cdot10^{-3}\,a_0^{-1}$ (
the other one has a much larger negative energy and is not represented in Fig.~\ref{figEV}, as it does not affect the low-energy scattering properties).

The behavior of the curves in Fig.~\ref{figEV}
is very similar to those of Fig.~\ref{figE},
in particular regarding the transformation of the Feshbach resonance into a virtual state.
The only difference between the present case (avoided crossing between a virtual state and a Feshbach resonance) and the previous case (avoided crossing between a bound state and a Feshbach resonance) is that here a virtual state transforms into a bound state before the crossing, while there a virtual state transforms into a bound state after the crossing.
Another interesting comparison is between our Fig.~\ref{figEV} and Fig.~5 of Ref.~\cite{marcelis:04};
it would be instructive to perform a detailed comparison of the two models to explain the differences between these two figures.

As for the interplay with a bound state, Fig.~\ref{figEV} also shows some limit on the range of magnetic field on which our model can be used:
since $\kappa_1$ is fixed in Eq.~\eqref{kappa1Rb} and the bound-state energy should be larger than $-\kappa_1^2 \approx -0.0075 a_0^{-2}$ (otherwise the potential becomes singular for some value of $r$),
the field should be lower than 24.5 mT.

\begin{widetext}
\begin{figure}
\begin{center}
\begin{minipage}{17cm}
\epsfig{file=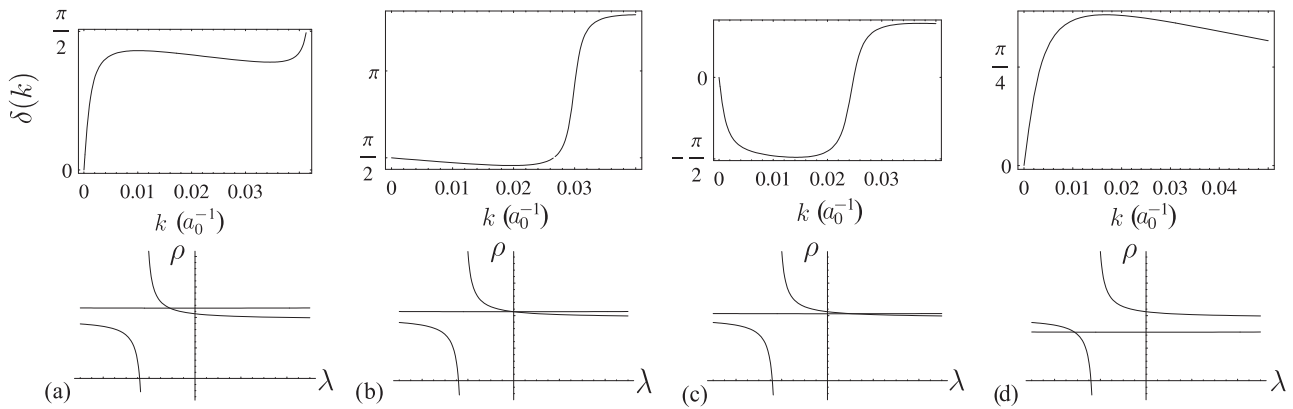, width=17cm} \caption{\small Phase
shifts and graphical representation of Eqs.~\eqref{syslr} for the Cox potential defined by parameters~\eqref{parRb}.
The columns correspond to different values of the magnetic field:
(a)~$B=$~14.454~mT; (b)~15.504~mT; (c)~15.854~mT ; (d)~19.0~mT.
\label{figSH1}}
\end{minipage}
\end{center}
\end{figure}
\end{widetext}

\begin{figure}
\begin{center}
\epsfig{file=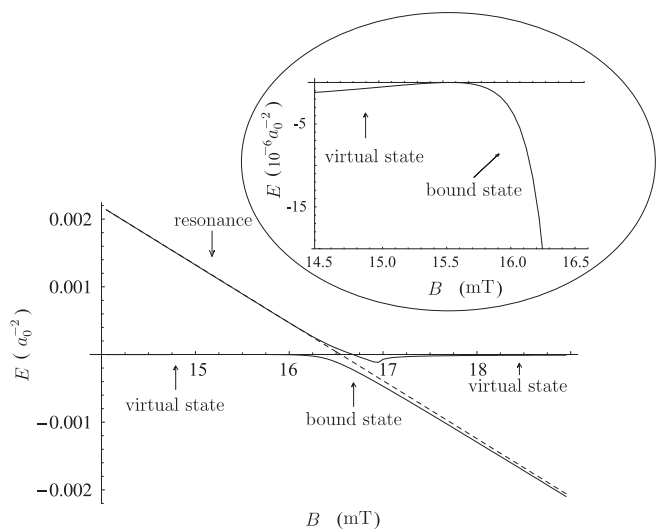, width=8.6cm}
\caption{\small $B$-dependence of the energies of the bare (dashed lines) and dressed (solid lines)
 states for the Cox potential defined by parameters~\eqref{parRb}.
\label{figEV}}
\end{center}
\end{figure}

The behavior of the phase shifts in the region
with the resonant and virtual states is shown in the first row of Fig.~\ref{figSH1}.
A similar discussion to that of Fig.~\ref{figSH} can be made here, except that here the large positive
 background scattering length results in a large negative slope for the phase shift at the origin.
Exactly at $B_0=15.5041$~mT, when the bound state transforms into a virtual state,
the phase shift starts from $\pi/2$.
The second row of Fig.~\ref{figSH1} shows the corresponding behaviour of the bound- and virtual-state zeros on the wave-number imaginary axes,
 confirming the above analysis.

\section{\label{sec:conclusion} Conclusion}

In this work, we have derived the exactly-solvable $N$-channel Cox potential
from a supersymmetric transformation of the vanishing potential and we have established
different parameterizations of this potential, as well as a necessary and sufficient condition for its regularity.
In the $N=2$ case, a full analysis of the corresponding Jost matrix has been carried out.
The structure of the zeros of the Jost determinant has been presented geometrically and
a method for controlling the position of the zeros of this Jost determinant has been proposed.
This has led to several examples of Cox potentials with different number of bound states and resonance,
solving schematic coupled-channel inverse problems.

With ultracold gases in mind, we have also studied the low energy $S$-matrix and the scattering length of the Cox potential.
Using independence of scattering properties from interaction details in the regime with a large scattering length,
a model of alkali-metal atom-atom scattering has been constructed.
This provides interesting exactly-solvable schematic models for the
interplay of a magnetically-induced Feshbach resonance with a bound state or a virtual state close to threshold.

We consider the development of supersymmetric transformations as a very promising tool for the multi-channel
inverse scattering problem and for the construction of more advanced exactly-solvable coupled-channel models.
In particular, iterations or chains of transformations might lead to more complicated Jost functions,
with arbitrary number of bound states and resonances, hopefully still with
a tractable connection between potential parameters and physical observables.

As far as physical applications are concerned, atom-atom interactions are both very interesting today, due to the active research field of ultracold gases, and rather simple with respect to supersymmetric quantum mechanics, as only $s$-waves have to be considered and as the interaction is short ranged (no Coulomb term).
We foresee to apply the present model to other systems presenting these simple features, namely coupled $s$-wave baryon-baryon interactions, with at least one neutral baryon.
In the longer term, we hope to generalize our method to higher partial waves and to Coulomb interactions. This should allow us to construct useful models in the context of low-energy nuclear reactions, the field which first motivated the work of Feshbach~\cite{feshbach:58,feshbach:62} on coupled-channel resonances, leading to possible applications in nuclear astrophysics and exotic-nuclei low-energy reactions.

\acknowledgments{We thank Daniel Baye for very useful discussions at several stages of this work and for drawing our attention to Ref.~\cite{nygaard:06}.
AP is supported by Russian "Dynasty" foundation.
BFS is partially supported by grant RFBR-06-02-16719.
AP and BFS are partially supported by grant SS-5103.2006.2;
they also thank the National Fund for Scientific Research, Belgium,
for support during their stays in Brussels.
This text presents research results of the Belgian program P6/23 on interuniversity
attraction poles of the Belgian Federal Science Policy Office (BriX, Belgian Research Initiative on eXotic nuclei).
}

\bibliographystyle{apsrev}
\bibliography{$HOME/Biblio/own,$HOME/Biblio/others}

\end{document}